\documentclass[fleqn,usenatbib]{mnras}
\usepackage[T1]{fontenc}
\usepackage{ae,aecompl}
\usepackage{xtab}
\usepackage{caption}
\usepackage{courier}
\usepackage{multirow}
\usepackage{times}
\usepackage{upgreek}
\usepackage{graphicx}
\usepackage{epstopdf}
\usepackage{amsmath}
\usepackage[section]{placeins}
\usepackage{tablefootnote}
\usepackage{color}
\usepackage{gensymb}

\newcommand{\gaia}{{\em Gaia}}
\newcommand{\kep}{{\em Kepler}}
\newcommand{\tess}{{\em TESS}}

\newcommand{\aador}{AA\,Dor}
\newcommand{\hwvir}{HW\,Vir}
\newcommand{\hwvirs}{HW-Vir systems}
\newcommand{\toby}{EQ\,1938+4603}
\newcommand{\romer}{R{\o}mer}
\newcommand{\msun}{M$_{\odot}$}

\newcommand{\teff}{\ensuremath{T_{\rm{eff}}}}
\newcommand{\logg}{\ensuremath{\log g}}

\usepackage{newtxtext}
\usepackage{newtxmath}

\begin{document}

\title[Space observations of \aador]{Space observations of AA\,Doradus provide consistent mass determinations. New HW-Vir systems observed with \tess.}

\author[A.S.\,Baran et al.]
{A.S.\,Baran,$^{1,2,3}$\thanks{E-mail: andrzej.baran@up.krakow.pl}
R.H.\,{\O}stensen,$^{1,2,4}$
U.\,Heber,$^{5}$
A.\,Irrgang,$^{5}$
S.\,Sanjayan,$^{1,6}$
J.H.\,Telting,$^{1,7,8}$
\newauthor
M.D.\,Reed$^{1,2}$
and
J.\,Ostrowski,$^{1}$
\\
$^{1}$ARDASTELLA Research Group, Institute of Physics, Pedagogical University of Krakow, ul. Podchor\c{a}\.zych 2, 30-084 Krak\'ow, Poland\\
$^{2}$Department of Physics, Astronomy, and Materials Science, Missouri State University, Springfield, MO\,65897, USA\\
$^{3}$Embry-Riddle Aeronautical University, Department of Physical Science, Daytona Beach, FL\,32114, USA\\
$^{4}$Recogito AS, Storgaten 72, N-8200 Fauske, Norway\\
$^{5}$Dr.\,Karl\,Remeis-Observatory \& ECAP, Astronomical Institute, Friedrich-Alexander University Erlangen-Nuremberg (FAU), Sternwartstr.\,7, 96049 Bamberg, Germany\\
$^{6}$Centrum Astronomiczne im. Miko{\l}aja Kopernika, Polskiej Akademii Nauk, ul. Bartycka 18, 00-716 Warszawa, Polska\\
$^{7}$Nordic Optical Telescope, Rambla Jos{\'e} Ana Fern{\'a}ndez P{\'e}rez 7, 38711 Bre{\~n}a Baja, Spain\\
$^{8}$Department of Physics and Astronomy, Aarhus University, NyMunkegade 120, DK-8000 Aarhus C, Denmark
}

\date{Accepted XXX. Received YYY; in original form ZZZ}
\pubyear{2019}

\label{firstpage}
\pagerange{\pageref{firstpage}--\pageref{lastpage}}
\maketitle

\begin{abstract}
We present an overview of eclipsing systems of the HW-Virginis type, based on space observations from the \tess\ Mission. We perform a detailed analysis of the properties of \aador, which was monitored for almost a full year. This excellent time-series dataset permitted us to search for both stellar pulsations and eclipse timing variations. In addition, we used the high-precision trigonometric parallax from \gaia\ Early Data Release 3 to make an independent determination of the fundamental stellar parameters. No convincing pulsations were detected down to a limit of 76 parts per million, however we detected one peak with false alarm probability of 0.2\%. 20\,sec cadences being collected during Year\,3 should confirm or reject our detection. From eclipse timing measurements we were able to confirm that the orbital period is stable, with an upper limit to any period change of 5.75$\cdot$10$^{-13}$\,s/s. The apparent offset of the secondary eclipse is consistent with the predicted \romer\ delay when the primary mass is that of a canonical extended horizontal branch star. Using parallax and a spectral energy distribution corroborates that the mass of the primary in \aador\ is canonical, and its radius and luminosity is consistent with an evolutionary state beyond core helium burning. The mass of the secondary is found to be at the limit of hydrogen burning.
\end{abstract}


\section{Introduction}
\label{intro}
Subdwarf B (sdB) stars are identified as compact stars located on the blue extension of the horizontal branch (EHB). The progenitors of sdB stars are intermediate-mass stars like the Sun that must have lost significant mass during or immediately after their ascent of the red-giant branch, leaving them with only a tiny remnant of their hydrogen envelopes. The mass loss must happen before helium ignition, otherwise they would become normal horizontal branch stars. Binary population synthesis modelling have been performed exploring various mass-loss scenarios, as detailed by \cite{han02}. Several channels exist, depending on the initial configuration of the system, and depending on the mass ratio, the binary system ends up either in a wide orbit (via stable Roche-lobe overflow, when the companion is sufficiently massive) or a close orbit (after common-envelope ejection, when the companion is of lower mass than the stripped EHB star). Since most EHB stars have a mass close to the core-helium flash mass of $\approx$ 0.5\,M$_{\odot}$ \citep{heber16}, the close-orbit systems can only consist of a sdB with either an M dwarf (dM), brown dwarf, or white dwarf companion.

\hwvir\ is the class prototype for eclipsing sdB+dM systems \citep{menzies86}. \cite{wolz18} provided a list of all \hwvirs\ studied prior to 2018. It contains 20 systems including those with brown dwarfs as secondaries. A large number of faint HW-Vir candidates from ground-based photometric surveys was recently published by \cite{schaffenroth19}. A typical light curve of \hwvirs\ shows two distinct eclipses and an out-of-eclipse variation, explained by an irradiation effect. The eclipses indicate a nearly edge-on orbital orientation (inclination close to 90$^{\circ}$). Eclipse mid-times can be used to study the stability of the orbital period, sometimes leading to the discovery of periodic modulations in the eclipse timings, which is indicative of additional companions \citep[e.g.][]{baran15a}.

\hwvirs\ are important objects for testing the proposed evolutionary channels of sdB stars described by \cite{han02}. Deriving the masses of both components is therefore crucial, but difficult. \hwvirs\ are usually single-lined spectroscopic binaries, since the secondary companions are not easily detectable in the presence of the more luminous hot stars. The luminosity ratio is of the order of 10$^6$. Masses of sdB stars are frequently assumed to be close to 0.47\msun~\citep{fontaine12}, which is often referred to as the canonical mass. In fact, a narrow range of helium-core masses is typical for a wide mass range of progenitors (0.7--1.9\msun) that undergo a helium flash. The canonical mass of about 0.47\msun\ is the flash mass for solar metallicity. With a wide range of metallicities, the permitted range for the core-helium flash extends to between 0.39 and 0.5 \msun~\citep{dorman93, han03}. However, a sdB can also have a mass of 0.31\,M$_{\odot}$ and up, if evolved of more massive progenitors that ignite helium under non-degenerate conditions \citep{1971ApJ...163..653H,han02,han03,2008A&A...490..243H,2017ASPC..509...85H,2020arXiv201114621O}. Such objects must be rare, though, because their progeny, low mass (0.3\,--\,0.45\,M$_\odot$) white dwarfs, are also rare \citep[for discussion see][]{heber16}. Thus, for any particular HW-Vir-type system, while a canonical mass for the primary is most likely, it is inappropriate to make this general assumption as it may lead to incorrect conclusions about other system parameters.

Stellar oscillations predicted and discovered in sdB stars by \cite{charpinet97} and \cite{kilkenny97}, respectively, are potentially useful for sdB mass estimations. Asteroseismology uses pulsation properties to describe stellar interiors, with a recent example of a mass estimation in EC\,21494-7018 reported by \cite{charpinet19}. Surprisingly, the estimated mass is 0.39\msun, which is significantly below the canonical mass. Other derived masses have been closer to the canonical one, \citep[see for example][]{charpinet11,baran19}. To date, the sample of pulsating sdB stars is approaching a hundred, with half of the sample discovered during the space photometry missions {\it Kepler}, {\it K2} and {\it TESS}.

Binary systems provide an independent tool to estimate masses. In the case of \hwvirs, the masses are not easily derived, since a significant temperature difference between components makes it difficult to see the secondary/fainter companion and typically the mass estimation for the secondary is based on an assumption of the primary having the canonical mass, which may not always be correct. \cite{kaplan10} reported of a new tool to estimate masses in binary systems. He showed that, even in the case of a circular orbit, a secondary eclipse is not centered at 0.5 orbital phase, but is observed with a lag, also known as the \romer\ delay. The predicted delay may be just a few seconds, and therefore very precise data are required. \cite{barlow12} measured that shift in \toby, one of the systems observed by the \kep\ spacecraft. They measured the delay to be below 2\,s and consequently the mass of the primary to be 0.37\msun. Next, \cite{baran15a} used a longer time span and arrived at a slightly smaller shift. These authors reported the mass of the primary to be smaller than 0.3\msun, which is contrary to the canonical mass. However, both \cite{barlow12} and \cite{baran15a} noted that the eccentricity of the orbit may also contribute to the offset of the secondary eclipse and the overall shift may not be purely caused by the \romer\ delay. In that case, the mass estimation may not be correct. \cite{baran18} reported an attempt to apply the \romer-delay method to \hwvir. The mass the authors derived was similar to that of \toby. They also arrived at the same conclusion about eccentricity, but noted that when the radii of the stars are large in relation to the orbital separation, several geometric effects can contribute to reduce the observed \romer\ delay. Together, these cases showed that the idea of \cite{kaplan10} may be tough to employ conclusively to \hwvirs, especially since it is impossible to observe eccentricity to the precision required for reasonably precise mass derivations. A mass as low as 0.3\msun\ can only be reconciled with evolutionary models if the hot subdwarf fails to ignite helium and evolves from the RGB directly to the white dwarf cooling track, in what is known as post-RGB systems.

\toby\ and \hwvir\ were observed during the \kep\ mission, which provided continuous time-series data of unprecedented quality. These were good enough to estimate the shift of the secondary eclipse. Another HW-Vir-like system observed during a space mission is \aador, which is located in the southern continuous viewing zone of the \tess\ satellite. A flux variation of \aador\ was first reported by \cite{kilkenny75} and this system has since been studied regularly. The light curve of \aador\ resembles that of \hwvir\, though because the estimated mass of the fainter component in \aador\ falls below the hydrogen core burning stellar configuration, the secondary was considered to be a brown dwarf. The relatively high effective temperature and low surface gravity places the \aador\ primary above the regular EHB region in the (\teff, \logg) diagram, which means that if its mass is canonical, it must have evolved beyond the core-helium-burning stage to the post-EHB shell-helium-burning stage. In this respect it is similar to V1828\,Aql \citep[=\,NSVS\,14256825,][]{almeida12} and EPIC\,216747137 \citep{silvotti21}.

To determine stellar masses reliably one needs to detect spectral lines of both components. A spectroscopic effort to determine the nature of the secondary component was undertaken {\it e.g.} by \cite{hilditch96} and \cite{rauch03}. No definite conclusion has been achieved, since the mass of the primary had to be assumed upfront. \cite{vuckovic08} re-analyzed the spectroscopic data of \cite{rauch03}, discovering emission lines originating from the heated side of the secondary. The authors made the first estimate of the masses of both \aador\ components, which were consistent with a regular EHB primary and a low-mass M dwarf secondary. A subsequent effort by \cite{2011A&A...531L...7K} ruled out the post-RGB channel. Updated work by \cite{hoyer15} and \cite{vuckovic16} produced the best estimations of radial velocity amplitudes of both components, allowing for precise mass determinations of both components. Both authors cited radial velocity amplitudes and corresponding masses that agree, within the errors, though the uncertainty of the radial velocity amplitude of the secondary component reported by \cite{vuckovic16} is an order of magnitude smaller. The masses indicate that the primary star is close to canonical, and the secondary is on the limit of hydrogen burning for a main sequence star.

In this paper we describe the \tess\ data in Section\,\ref{sect:photdata}, search for pulsations in Section\,\ref{sect:puls} and explore eclipse timings in Section\,\ref{sect:midtimes}. We report results of our work using \tess\ data of \aador\ in Section\,\ref{sect:roemer}, in which we estimate masses of both components. In Section\,\ref{sect:stellar_param} we explore another method to derive the fundamental stellar parameters using published atmospheric parameters, modelling the spectral energy distribution and making use of the high-precision trigonometric parallax provided by the \gaia\ Early Data Release 3 \citep[EDR3,][]{2016A&A...595A...1G,2020arXiv201201533G}. Finally, in Section\,\ref{sect:other_systems} we present an overview of \tess\ systems of other \hwvirs, and discuss the potential of mass determinations for these.

\begin{figure}
\includegraphics[width=\hsize]{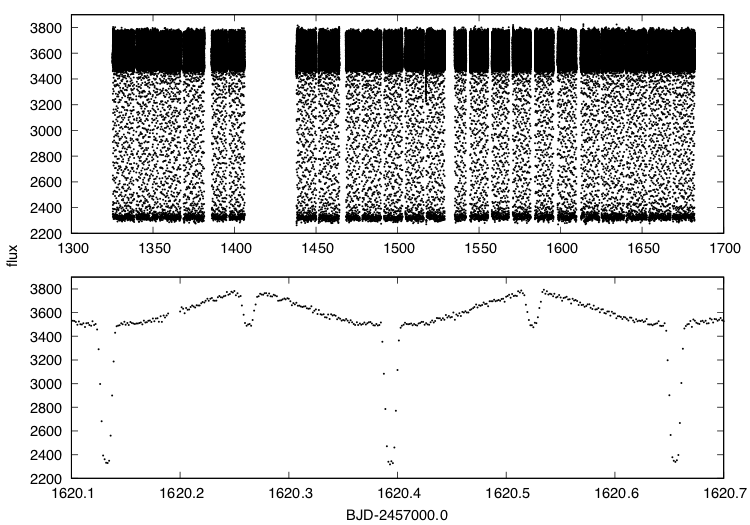}
\caption{Light curve of \aador\ obtained by \tess\ during Year 1 of the mission. The top panel shows the entire processed data set, while the bottom panel shows the flux variations during a couple of orbital periods.}
\label{fig:lc}
\end{figure}

\section{TESS photometric data}
\label{sect:photdata}
\aador\ ($\alpha_{2000}=05^{\rm h}31^{\rm m}40.36^{\rm s}$, $\delta_{2000}=-69^{\circ}53'02.2''$) was observed during the first cycle of the nearly all-sky survey undertaken with the Transiting Exoplanet Survey Satellite (\tess). \tess\ is deployed in an elliptical, 2:1 lunar synchronous orbit with a period of 13.7\,d. Each annual cycle of \tess\ observations are split up into sectors lasting two orbits, or about 27\,d. The detector consists of four contiguous CCD cameras, each covering a 24$^\circ$\,x\,24$^\circ$ field of view (FoV), making up a 24$^\circ$\,x\,96$^\circ$ strip aligned along ecliptic latitude lines. The data are stored with the short cadence (SC), lasting 120\,s and the long cadence (LC), lasting 1800\,s. When one sector observations have been completed, the instrument's FoV is shifted eastward by 27$^\circ$, naturally pivoting around the ecliptic pole. It takes 13 sectors to pivot around one pole, then the FoV is shifted to the other hemisphere for the next cycle. As a result, the regions near the ecliptic poles are observed during every sector and are known as the continuous viewing zones of \tess. Luckily, \aador\ is located in this zone around the southern ecliptic pole. We downloaded all available data from the ``Barbara A. Mikulski Archive for Space Telescopes'' (MAST)\footnote{archive.stsci.edu}. The data span all 13 sectors with the exception of Sector 4, during which \aador\ was not included in the target list. We used the SC data which has a time resolution high enough to allow us to sample the eclipses and to search for stellar pulsations up to 4166\,$\upmu$Hz, covering both the g-mode and partially the p-mode regions in an amplitude spectrum. We extracted PDCSAP\_FLUX, which is corrected for on-board systematics and neighbors' contribution to the overall flux. We clipped fluxes at 5$\upsigma$ to remove outliers, de-trended long term variations (of the order of days) with polynomials. We show the resultant light curve in Figure\,\ref{fig:lc}.

\begin{figure}
\centering
\includegraphics[width=\hsize]{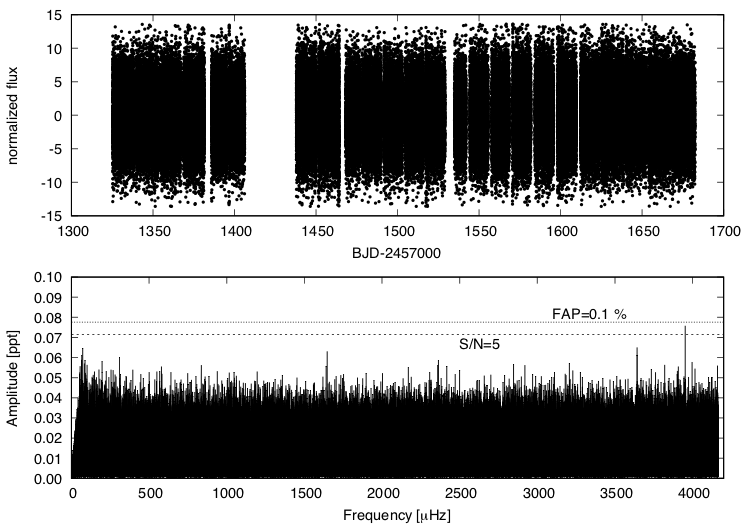}
\caption{Top panel: residuals of prewhitening and clipping. Bottom panel: amplitude spectrum up to the Nyquist frequency (4166\,$\upmu$Hz). The dashed line in the bottom panel represents S/N\,=\,5 (0.0714\,ppt). The dotted line represents FAP\,=\,0.1\%.}
\label{fig:puls}
\end{figure}

\section{Stellar pulsations}
\label{sect:puls}
A few of the sdB primaries in \hwvirs\ show stellar pulsations. The examples are NY\,Vir, \toby\ and \hwvir. Therefore, we made an attempt to detect stellar pulsations in \aador. The light curve of \aador\ is dominated by eclipses and an irradiation effect. In order to detect any pulsations, which would typically have amplitudes close to the 10 parts per thousand (ppt) level, we had to remove the binary orbital signature from the data. First, we used a Fourier domain to calculate the orbital frequency and a sequence of harmonics, which appear as a consequence of a non-sinusoidal shape of the flux variation. Then, we prewhitened the binary frequency along with 93 harmonics, and finally clipped the residuals to remove data points that became outliers after prewhitening. We present the result of these processed data in Figure\,\ref{fig:puls}. The light curve no longer shows any binary trend. The amplitude spectrum is fairly smooth with just a few low amplitude frequencies. The highest amplitude peak is at 3952.19\,$\upmu$Hz with an amplitude of 0.076\,ppt. The signal-to-noise (S/N) of this peak exceeds the S/N\,=\,5 criterion, where N is a median noise level. To determine the significance of the peak, we simulated 10\,000 pure-noise time series data sets sampled exactly as the \tess\ data. A false-alarm-probability (FAP) of 0.1\% is achieved at 0.0776\,ppt level while the FAP at 0.076\,ppt is 0.2\%. The peak at 3952.19\,$\upmu$Hz may be intrinsic to \aador, however it is close to the Nyquist frequency so its true frequency may be 4381.14\,$\upmu$Hz. The 20\,sec cadences will be of high importance to confirm this detection. The amplitude of the peak is barely meeting our FAP\,=\,0.1\% threshold, so there is still possibility, though not too high, that the peak is of noise origin randomly bumped up to to that level.

\begin{figure}
\centering
\includegraphics[width=\hsize]{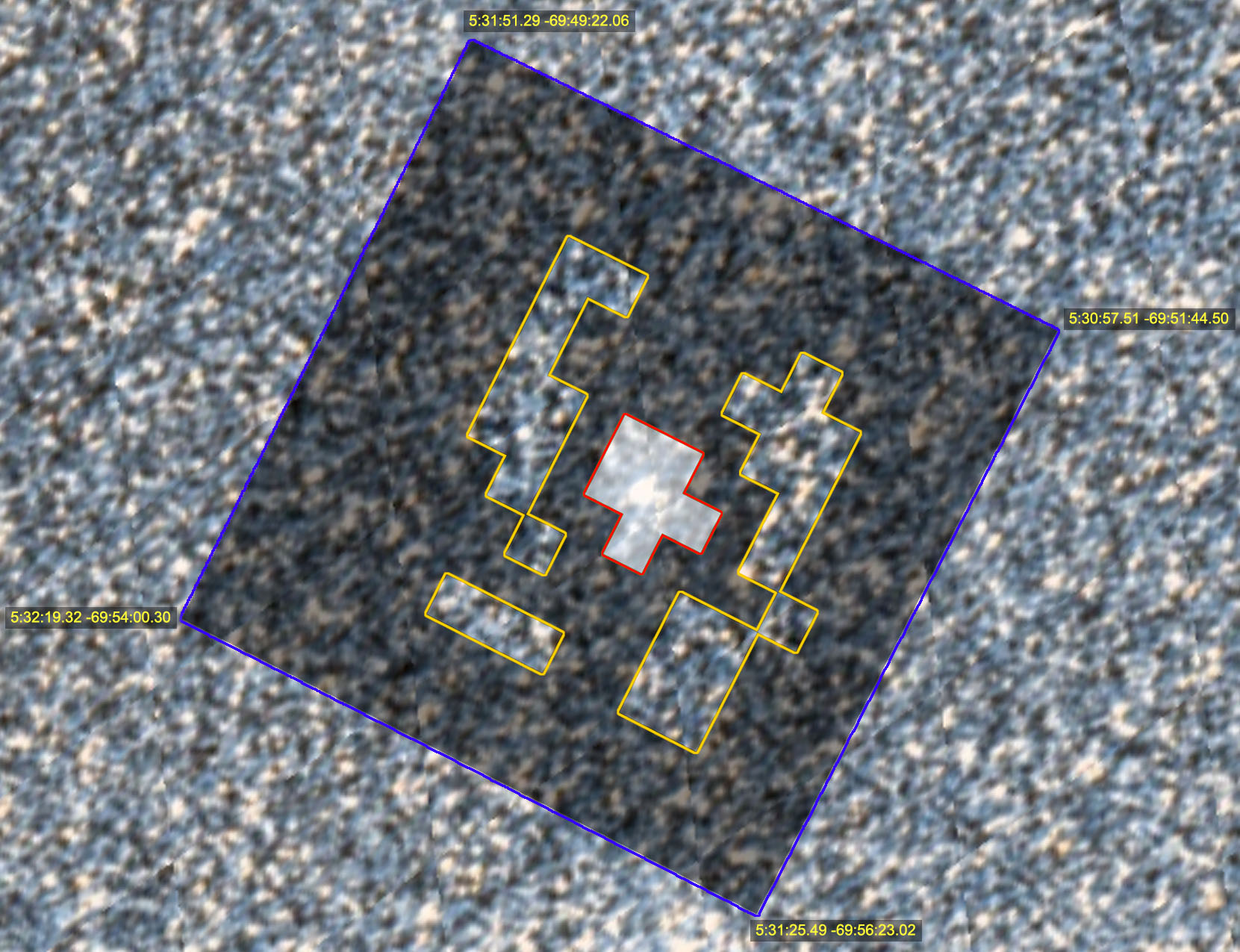}
\caption{Image showing the part of the sky around \aador. The red aperture is used for the target flux, while the yellow one is used for the sky background estimation.}
\label{fig:image}
\end{figure}

With an effective temperature of 42,000\,K \citep{2000A&A...356..665R}, the \aador\ primary is on the hot end of known sdB pulsators, where most sdB stars are non-pulsators. Exceptions have been the sdO \citep{woudt06} and $\omega$\,Cen pulsators \citep{brown13}. Temperatures between 40,000 and 50,000\,K may not allow for observable pulsation amplitudes with currently available time-series data sets. Another reason may be that the amplitudes are diluted by an overestimated flux level. The pulsation amplitudes in other HW-Vir-system primaries, are around 0.1\,ppt. If the \aador\ primary has amplitudes below 0.1\,ppt, they will be diluted since \aador\ is located in front of a dense stellar environment, namely the Large Magellanic Cloud. In Figure\,\ref{fig:image} we show a part of the sky around \aador\ with the target mask and the optimal aperture overplotted. This figure shows that the CCD pixels used in the optimal aperture for the target overestimate \aador\;'s flux, (according to the crowding metric, by 23\%, on average), while those used for sky background overestimate the true sky flux. This leads to a dilution, (by 23\%, on average), in amplitude of any flux variations intrinsic to \aador.

\section{The mid-times of eclipses}
\label{sect:midtimes}
\begin{figure*}
\centering
\includegraphics[width=\hsize]{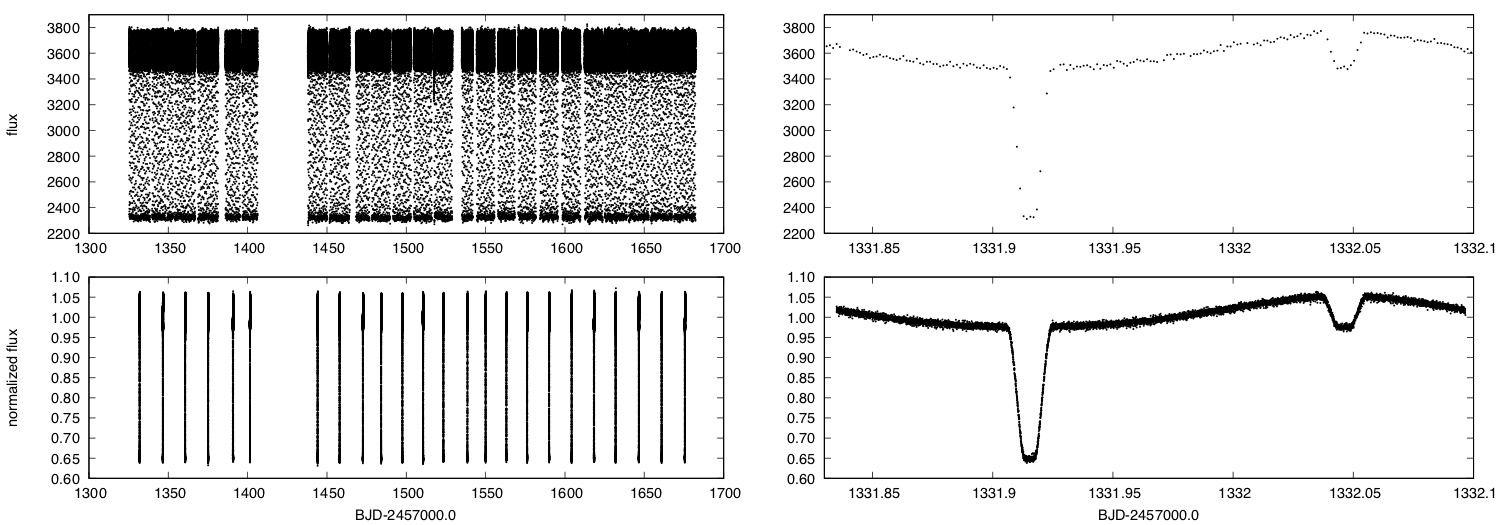}
\caption{Left upper panel shows the original light curve, while left bottom one shows the light curve folded over the binary orbital period during each \tess\ orbit. The upper right panel shows an unfolded binary orbit, while the bottom right one shows a folded one.}
\label{fig:meaneclipse}
\end{figure*}

Flux variations caused by orbital motion include two eclipses and an irradiation effect. The eclipses are used to study the stability of the orbital period. The mid-times of eclipses are plotted in the so-called {\it Observed minus Calculated} (O--C) diagram, which can be used to measure an orbital period variation \citep{sterken05}. Its variability provides clues on {\it e.g.} possible mass exchange, a tertiary body, or gravitational wave radiation. To calculate mid-times we used the method described in \cite{kwee56}. Since we detected no significant pulsations, eclipse shapes are not distorted by other variability and are strictly defined by the geometry of the system. However, to increase the sampling during eclipses and to lower uncertainties of the mid-times, we decided to fold each part of the light curve obtained during a single \tess\ orbit over the binary period calculated from all-sector individual eclipses. This folding increases precision of data in an O--C diagram with the price of decreased time resolution, since the total number of eclipses is decreased to 24. However, we do not expect any orbital period variations on time scales of days or less. We show the result of our light curve folding in Figure\,\ref{fig:meaneclipse}.

Having the light curve folded, we recalculated the ephemeris from the mid-times of the primary eclipses and obtained
\[ 
{\rm T_{pri}}=2455681.9153693\,(32)\,{\rm BJD}+0.2615397323\,(40)\,{\rm E}
\]

Likewise, we re-calculated the ephemeris from the mid-times of the secondary eclipses and obtained
\[ 
{\rm T_{sec}}=2455682.046202\,(9)\,{\rm BJD}+0.261539724\,(11)\,{\rm E}
\]

\begin{figure}
\includegraphics[width=\hsize]{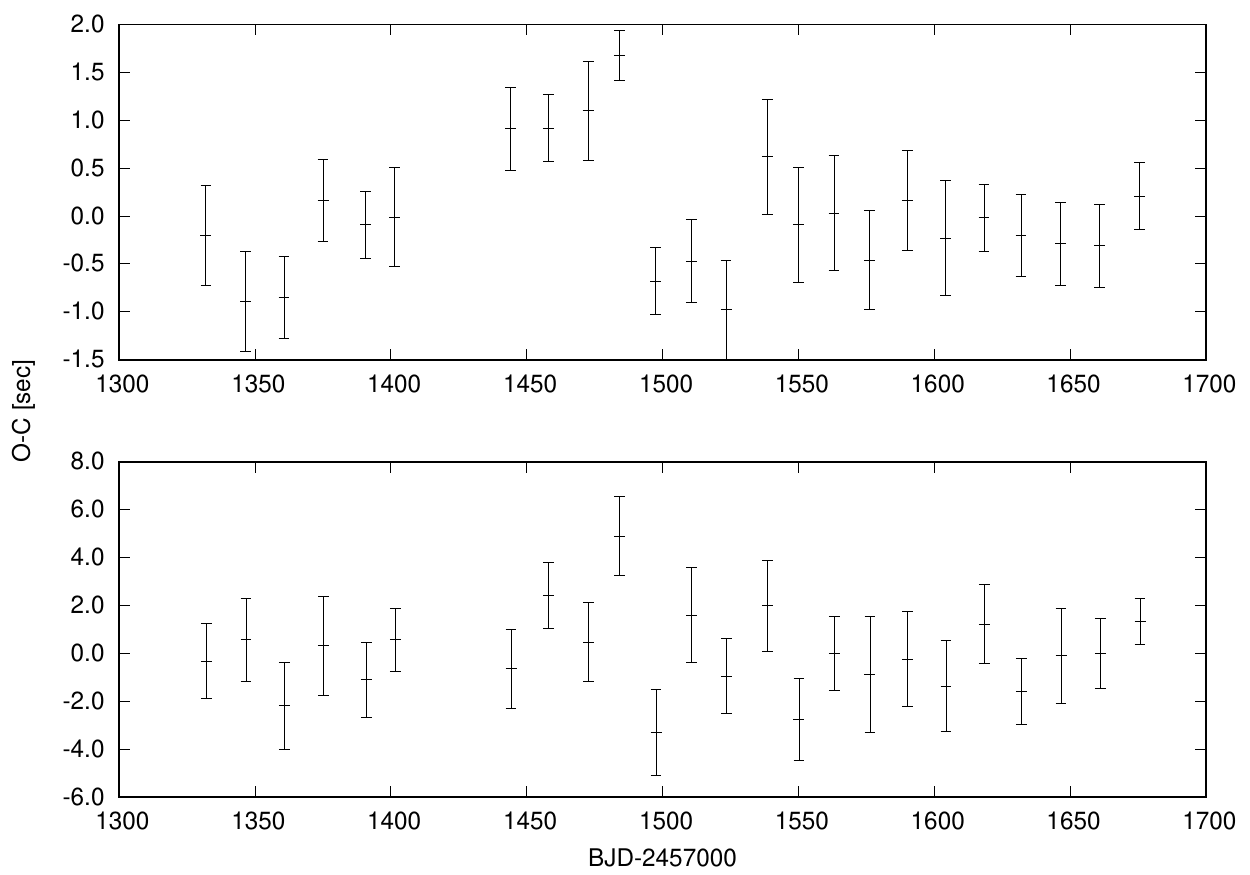}
\caption{The O-C diagrams for primary (upper panel) and secondary (bottom panel) eclipses.}
\label{fig:oc}
\end{figure}

The orbital period derived from both types of eclipses agrees very well to within the errors. We plot the O-C diagrams for primary and secondary eclipses in Figure\,\ref{fig:oc}.

The orbital period has been monitored for almost half a century. In Table\,\ref{tab:pastperiod} we list published values of the orbital period. Dave Kilkenny of the South African Astronomical Observatory and his collaborators have been very active in this field. The consecutive values of the orbital period are derived from all mid-times available to the authors. We can see how a longer time baseline helps increase the precision of the orbital period. \cite{kilkenny11} presented the most precise orbital period and the O--C analysis, which shows that the orbital period remains very stable over 30\,years of monitoring (their Figures 1 and 2). Our work confirms that the orbital period remains stable. The period difference between 1981 and 2019 is 0.69\,msec, which gives an upper limit for period change to be 5.75$\cdot$10$^{-13}$\,s/s. The O--C diagram does not indicate any tertiary body in the system, while other effects, which could cause period change, must be negligible.

\begin{table}
\caption{The orbital period of \aador\ derived since its discovery. The last listed value, generously provided by Dave Kilkenny, has been calculated from eclipses collected between February 1977 and March 2020.}
\label{tab:pastperiod}
\centering
\scalebox{1.0}
{
\begin{tabular}{llc}
\hline\hline
Period [days] & Uncertainty & Reference \\
\hline\hline
0.261539 & 0.17\,sec & \cite{kilkenny78} \\
0.2615398 & 17\,msec & \cite{kilkenny79} \\
0.261539724 & 0.35\,msec & \cite{kilkenny83} \\
0.261539726 & 0.26\,msec & \cite{kilkenny86} \\
0.2615397198 & 0.15\,msec & \cite{kilkenny91} \\
0.261539731 & 0.17\,msec & \cite{kilkenny00} \\
0.2615397363 & 35\,$\upmu$sec & \cite{kilkenny11} \\
0.2615397323 & 0.35\,msec & this work \\
0.2615397364 & 35\,$\upmu$sec & D.Kilkenny (priv. comm.) \\
\hline\hline
\end{tabular}
}
\end{table}

\begin{figure}
\includegraphics[width=\hsize]{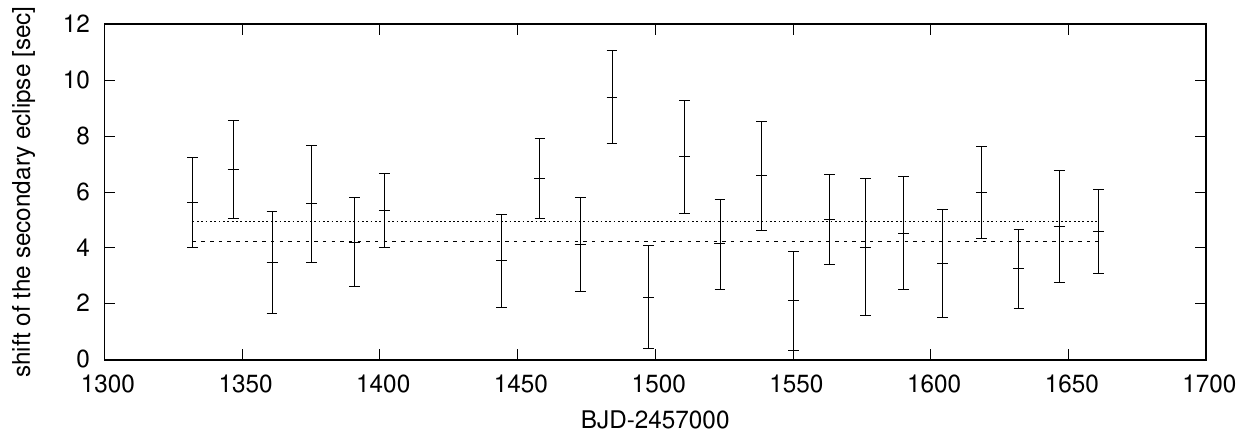}
\caption{The shift of the secondary eclipse measured in folded and binned eclipses. Two horizontal lines show the range of the mean value of the shift.}
\label{fig:romer}
\end{figure}

\section{Shift of the secondary eclipse} \label{sect:roemer}
Binary systems with components of unequal masses must have mid-times of secondary eclipses which happen slightly later than half an orbital phase after the primary eclipses. The effect is caused by the finite speed of light and is often called the \romer\ delay. This shift must be present unless the masses of both components are equal or the orbit has a specifically-tuned eccentricity which cancels out the shift. In the case of circular orbits, the offset of the secondary eclipse is purely due to the \romer\ delay, and gives a direct measure of the mass ratio, whenever the orbital velocity of either component and the orbital period are known. In this case, all these quantities can be measured, so the mass estimates are purely observational and do not require any modeling or calibrations. All necessary formulae can be found, {\it e.g.} in \cite{baran18}.

We calculated the shifts from 24 folded primary and secondary eclipses. Then we took a mean value, which we found to be 4.59\,(36)\,s. The uncertainty is defined as the error of the mean, so it is purely statistical error. We plotted our calculated shifts, along with the range of the mean value in Figure\,\ref{fig:romer}. Since \cite{hoyer15} and \cite{vuckovic16} reported the values of the radial velocity amplitudes of both components, we can calculate the expected shift of the secondary eclipses, assuming the orbit remains circular. \cite{hoyer15} cited a value of K$_2$ to be 232.9$^{16.6}_{-6.5}$\,km/s and K$_1$ to be 40.15\,(11)\,km/s. \cite{vuckovic16} cited K$_2$ to be 231.3\,(7)\,km/s and K$_1$ to be 39.63\,(21)\,km/s. The uncertainty of K$_2$ cited by \cite{vuckovic16} is smaller, therefore, for consistency, we adopted values of K$_1$ and K$_2$ values and consequently, the mass ratio, from \cite{vuckovic16}. We adopted the orbital period from our ephemeris, calculated from folded primary eclipses. Then, the expected shift of the secondary eclipses from pure \romer\ delay, {\it i.e.} assuming circular orbit, equals 4.599\,(28)\,s. This value agrees very well with the one we derived from our observations, measuring the shift directly from the secondary eclipses.

Using K1, P, i\,=\,90$^\circ$ and our measured \romer\ delay, we derive values for the masses of 0.46\,(5)\,M$_{\odot}$ and 0.079\,(9)\,M$_{\odot}$ for the primary and the secondary components, respectively. The masses are in agreement with those derived by both \cite{hoyer15} and \cite{vuckovic16} and confirm that the primary has a mass close to the canonical value. The \romer\ delay measurements are consistent with the orbit having e\,=\,0. If the orbit were eccentric, it would contribute to the shift of secondary eclipses according to Equation\,8 in \cite{baran18}. There are two parameters that are essential, eccentricity $e$ and the argument of pericentre, $\omega$. If the value of $e$ is non-zero and $\omega$ is different from $\pi$/2 or 3$\pi$/2, the shift we derived from eclipses would be bigger or smaller depending on the sign of $\cos\omega$. While we cannot completely rule out a non-zero eccentricity, if the orbit were eccentric, while $\cos\omega$ is close to zero, the contribution to the shift of the secondary eclipse would remain negligible, so the shift of the secondary agrees well with the \romer\ delay. Unfortunately, a small eccentricity is not detectable in current observations so a definite conclusion cannot be made. Perhaps, an apsidal motion caused by a precession of an eccentric orbit would help to set the limit on eccentricity, however observations do not indicate this to be the case. Convincing observations to measure precise eccentricity are yet to be collected.

\section{Stellar Parameters: Radius, mass, and luminosity}\label{sect:stellar_param}
A high precision measurement of the parallax of AA\,Dor 
($\varpi$\,=\,2.8380$\pm$0.0441\,mas) is available through the \gaia\ EDR3, which allows us to derive the stellar parameters radius, mass, and luminosity from the atmospheric parameters (\teff and \logg from spectroscopy). To this end the angular diameter is derived from the spectral energy distribution (SED), which combined with the parallax yields the stellar radius.    

\subsection{Angular diameter and interstellar reddening}

The angular diameter $\Theta$ is a scaling factor from $f(\lambda)=\Theta^2 F(\lambda)/4$, where $f(\lambda)$ and $F(\lambda)$ are the observed and synthetic stellar surface fluxes, respectively. Because of the light contribution originating from the heated hemisphere of the companion, the sdB fluxes can reliably be measured only when the companion is completely eclipsed by the larger subdwarf, that is at secondary eclipse \citep[see e.g.][]{2020MNRAS.tmp.3451S}
Such data are not available for \aador. Nevertheless, many photometric measurements are available in different filter systems, covering the optical (\gaia\ EDR3 \citep{2020arXiv201201916R}, APASS \citep{2015AAS...22533616H}, SkyMapper DR2 \citep{2019PASA...36...33O}, Tycho \citep{2000A&A...355L..27H}, 
and infrared (2MASS \citep{2003yCat.2246....0C}, DENIS \citep{2000A&AS..141..313F}, and WISE \citep{2019ApJS..240...30S}) spectral ranges.
Johnson and Str\"omgren colours are also available \citep{1998A&AS..129..431H,2003AJ....125.2531R}. 

Many observations of \aador\ in the ultraviolet were made with the {\it International Ultraviolet Explorer (IUE)}, which were retrieved from the Final Merged Log of IUE Observations through Vizier\footnote{\url{https://cdsarc.unistra.fr/viz-bin/cat/VI/110}} to derive UV-magnitudes from IUE spectra. Three box filters, which cover the spectral ranges 1300--1800\AA, 2000--2500\AA, and 2500--3000\AA\ are defined \citep{2018OAst...27...35H}. We averaged magnitudes derived from 13 SWP and 12 LWR low resolution (6\AA) spectra taken through the large (20\,arcsec) aperture.

Published photometric measurements are mostly averaged from observations taken at multiple epochs. Hence, they may be affected by some extra light from the companion.

Another factor that influences the spectral energy distribution is interstellar reddening. Even though interstellar extinction is probably low at high Galactic latitudes, it has to be taken into account. We use the reddening law of \citet{2019ApJ...886..108F} and the angular diameter is determined simultaneously with the interstellar colour excess. A $\chi^2$ based fitting routine is used to match  synthetic flux distributions from the grid of model atmospheres calculated with ALTLAS12 to the observed magnitudes \citep[see][for details] {2018OAst...27...35H}. We use spectroscopic parameters of \teff\,=\,42,000(1000)\,K and \logg=\,5.46(5) \citep{2011A&A...531L...7K} which were derived using metal line blanketed NLTE models calculated with the T\"ubingen model atmosphere code TMAP and metal abundances from detailed quantitative spectral analyses of UV spectra \citep{2008A&A...492..565F}. The helium abundance is n$_{\text He}$/n$_{\text H}$\,=\,0.0008(2) \citep{2000A&A...356..665R}.

\begin{figure}
\centering
\includegraphics[width=1\columnwidth]{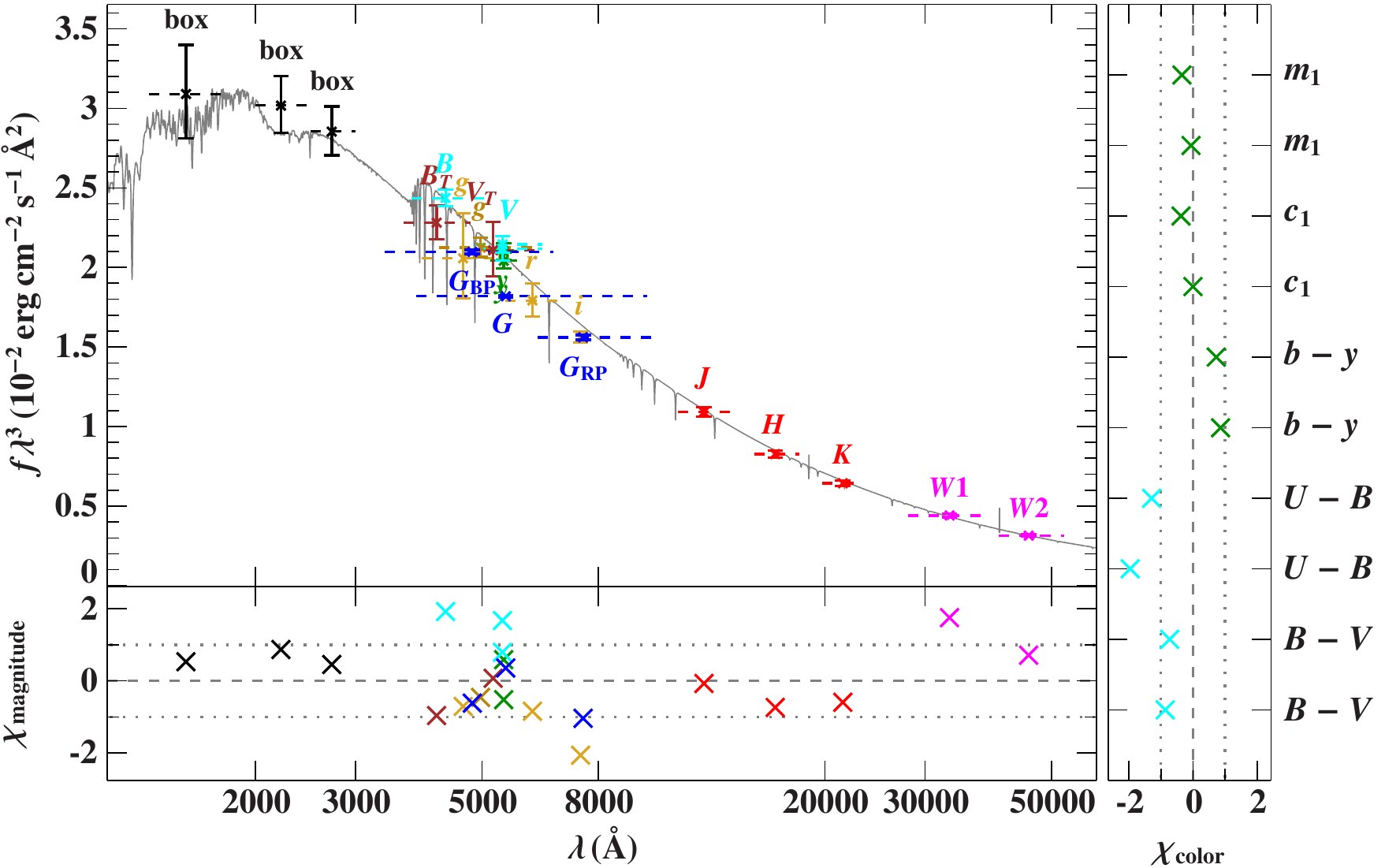}
\caption{\label{fig:photometry_sed_single} Comparison of synthetic and observed photometry (flux times wavelength to the power of three) and colours: \textit{Top panel, left hand side:} Spectral energy distribution of \aador\: Filter-averaged fluxes converted from observed magnitudes. Dashed horizontal lines depict the approximate width of the respective filters (widths at 10\% of maximum). The best-fitting model, smoothed to a spectral resolution of 6\,{\AA}, is shown in gray. \textit{Bottom panel:} Residual $\chi$: Difference between synthetic and observed magnitudes divided by the corresponding uncertainties. \textit{right hand panel:} Residual $\chi$ for Johnson and Str\"omgren colours. Difference between synthetic and observed colours divided by the corresponding uncertainties. The different photometric systems are displayed in the following colours:      (APASS-griz; 
SkyMapper DR2-g, golden);
Johnson (APASS, cyan);
Johnson (Tycho, brown);
\gaia\ (blue);
2MASS (red); 
DENIS-I (yellow); 
WISE (magenta). 
UV magnitudes derived from IUE spectra are labelled ''box'' (see text). Parameters resulting from the fit are listed in Table \ref{tab:sed_single}.}
\end{figure}

A photometric excess noise was derived in the fit procedure to ensure that the reduced $\chi^2$ is unity (see Table \ref{tab:sed_single} it was added to the uncertainties of all photometric magnitudes in quadrature. The most precise photometry comes from \gaia\ EDR3, based on 256 G band observations. While the \gaia\ EDR3 photometry dominates the determination of the angular diameter, additional optical and UV photometry is crucial to determine the interstellar reddening, and IR photometry could signal light from the companion via an IR excess. Judged by the very small photometric excess noise of 0.003mag, the match of the observed magnitudes is excellent (see Fig. \ref{fig:photometry_sed_single}). No infrared excess can be noticed. Numerical experiments adding a black body or cool star atmosphere failed to detect any extra flux contributions. We conclude that the light variations are averaged out by the large number of observations.

\begin{table}
\caption{Results for a single spectrum fit.}\label{tab:sed_single}
\begin{tabular}{lr}
\hline
\hline
Color excess $E(B-V)$ & $0.035\pm0.004$\,mag \\
Angular diameter $\log(\Theta\,\mathrm{(rad)})$ & $-10.570\pm0.007$ \\
Radius $R = \Theta/(2\varpi)$ & $0.209\pm0.005$\,$R_\odot$ \\
Mass $M = g R^2/G$ & $0.45\pm0.06$\,$M_\odot$ \\
Luminosity $L/L_\odot = (R/R_\odot)^2(T_\mathrm{eff}/T_{\mathrm{eff},\odot})^4$ & $122\pm13$ \\
\hline
Generic excess noise $\delta_\textnormal{excess}$ & $0.003$\,mag \\
\hline
\end{tabular}

\end{table}



\subsection{Stellar radius, mass and luminosity}
In order to determine stellar parameters we corrected
the Gaia EDR3 parallax of \aador\ for the zero point offset of -0.026 mas \citep[calculated using the prescription of][]{2020arXiv201203380L}.
From the atmospheric parameters (\logg and \teff) and the angular diameter the stellar radius $R$, mass $M$, and luminosity $L$ are derived. Their respective uncertainties are derived via Monte Carlo error propagation. 

The radius $R$ is derived from parallax and angular diameter. Then the mass follows from the spectroscopic gravity. Finally, the luminosity results from the radius and spectroscopic \teff. For the subdwarf, the SED fit gives a mass of $M\,=\,0.45\pm0.06\,M_\odot$, a radius of $R$\,=$0.209\pm0.005$\,$R_\odot$ and a luminosity of $L\,=\,122\pm13\,L_\odot$. The interstellar colour excess is small ($E(B-V)\,=0.035\pm0.004$\,mag).

It should be noted that the mass of the sdOB primary is consistent with the canonical evolutionary mass. Because of the high quality of the Gaia parallax, the sdOB radius is very precise and in good agreement with the result of the light curve analysis by \citet{2003MNRAS.344..644H}.

\begin{figure}
\includegraphics[width=\hsize]{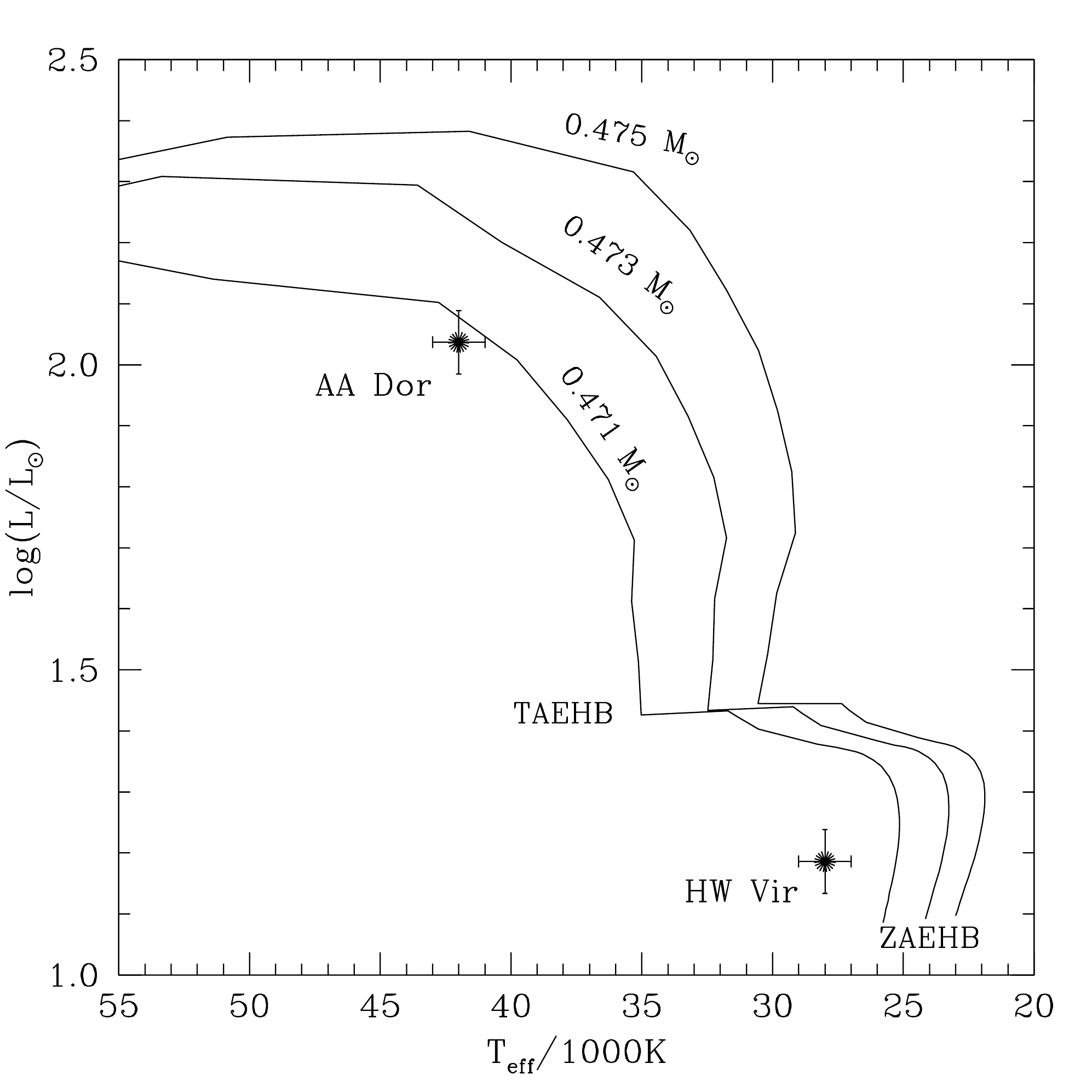}
\caption{Positions of \aador\ and \hwvir\ in the Hertzsprung-Russell diagram compared to evolutionary EHB tracks from \citet{dorman93} labeled by their stellar masses. The zero-age EHB is indicated as well as the terminal-age EHB (termination of core helium burning).}
\label{fig:hrd}
\end{figure}

Finally, we placed \aador\ in the Hertzsprung-Russell diagram (Figure\,\ref{fig:hrd}) to compare its position with canonical evolutionary models from \citet{dorman93}. For \hwvir, \citet{baran18} derived an astrometric radius from its Gaia parallax in a similar way as done here for \aador. Using the effective temperature of 28,000\,K \citep{vuckovic14} we derive its astrometric luminosity. The comparison in Figure\,\ref{fig:hrd} demonstrates that \aador\ is in an evolved state of evolution beyond core-helium exhaustion, while \hwvir\ is a bona-fide EHB star. The positions of both stars are consistent with evolutionary models of canonical mass. \citet{2000A&A...356..665R} suggested that \aador\ could be a stripped red giant branch star (post-RGB scenario), that is a star that left the RGB before helium ignition in the core and will evolve into a helium white dwarf. Appropriate evolutionary models predict that its mass should be 0.33 M$_\odot$ \citep{2000A&A...356..665R}. Because the mass of \aador\ derived here is significantly higher, the post-RGB scenario is ruled out.  

\begin{table*}
\caption{Known \hwvirs\ and new candidates observed with \tess. Only systems observed with either \tess\ or {\it Kepler} in SC mode have been included. Systems with detailed studies in the literature have the radial velocity amplitude, $K_1$ given. G-magnitudes and parallax from Gaia EDR3 have been included. The Sectors column provide the \tess\ sectors with SC observations, and Kepler observations are indicated with K1 for the main mission and K2 for the extended mission.}
\label{tab:hwvirs}
\centering
\begin{tabular}{lccccccll}
\hline
\multirow{2}{*}{Name} & \multirow{2}{*}{V* name} & \multirow{2}{*}{TIC} & P & $K_1$ & G & Parallax & \multirow{2}{*}{Sectors} & \multirow{2}{*}{Reference} \\
& & & d & km/s & mag & mas & \\
\hline
GALEX\,J194442.8+544942&  & 467187065 & 0.0642 &    - & 15.75 & 0.64(4) & S14,16 & \cite{schaffenroth19} \\
PG\,1621+476          &  & 193555713 & 0.0698 & 47.0 & 16.22 & 0.55(4) & S23-25 & \cite{schaffenroth14b} \\
KPD\,2045+5136        &  & 365213081 & 0.0896 &    - & 15.23 & 0.93(2) & S15,16  & \cite{schaffenroth19} \\
HE\,0516--2311        &  & 408187719 & 0.0912 &    - & 15.89 & 0.42(4) & S5,6    & \cite{schaffenroth19} \\
PTF1\,J011339.09+225739.1&&611402948 & 0.0934 & 74.2 & 16.61 & 0.42(7) & S17     & \cite{wolz18} \\
HS\,0705+6700& V470\,Cam &  99641129 & 0.0956 & 85.8 & 14.62 & 0.79(3) & S20     & \cite{drechsel01} \\
SDSS\,J0820+0008      &  & 455206965 & 0.0962 & 47.4 & 15.16 & 0.66(5) & S7      & \cite{geier11} \\
PG\,1336--018& NY\,Vir & 175402069 & 0.1010 & 78.6 & 13.37 & 1.68(4) & S23     & \cite{vuckovic07} \\
J19065+2807           &  & 281948821 & 0.1121 &    - & 15.64 & 0.64(3) & S14     & \cite{schaffenroth19} \\
BD--07$^\circ$3477&\hwvir& 156618553 & 0.1168 & 82.3 & 10.59 & 5.77(6) & K2      & \cite{baran18} \\
EC\,10246--2707       &  & 193092806 & 0.1185 & 71.6 & 14.44 & 1.02(4) & S9      & \cite{barlow13} \\
\toby       & Kepler-451 & 271164763 & 0.1258 & 65.7 & 12.11 & 2.44(3) & K1,S14,15  & \cite{ostensen10} \\
EVR-CB-003            &  & 396004353 & 0.1315 &    - & 13.51 & 1.77(2) & S11,12  & \cite{ratzloff20} \\
ASAS\,J102322--3737.0 &  &  73764693 & 0.1393 & 81.0 & 11.69 & 3.54(4) & S9      & \cite{schaffenroth13} \\
FBS\,1531+381         &  & 148785530 & 0.1618 & 71.1 & 12.94 & 1.90(3) & S24     & \cite{for10} \\
J21469+6616           &  & 322390461 & 0.1935 &    - & 16.21 & 0.68(3) & S15,18,24 & New \\
FBS\,0747+725         &  & 441613385 & 0.2083 &    - & 16.48 & 0.38(6) & S20,26  & \cite{pribulla13} \\
LB\,3459        & \aador & 425064757 & 0.2615 & 39.2 & 11.16 & 2.84(4) & S1-3,5-13 & \cite{vuckovic16} \\
EC\,23068--4801       &  & 139266474 & 0.2641 &    - & 15.36 & 0.70(4) & S1      & \cite{drake17} \\
Ton\,301              &  & 165797593 & 0.3697 &    - & 13.78 & 1.20(3) & S20     & \cite{schaffenroth19} \\
EC\,02406--6908       &  & 259864042 & 0.4607 &    - & 14.65 & 0.93(3) & S1,2    & New \\
\hline
\end{tabular}
\end{table*}

\section{HW-Vir systems observed with TESS}\label{sect:other_systems}

We explored the public \tess\ archive for HW-Vir systems by matching targets with the \cite{geier17} Gaia-selected hot-subdwarf sample. Table\,\ref{tab:hwvirs} summarises these results for the 21 systems that currently have space data with short-cadence sampling, and include eleven systems that are well-studied with spectroscopic orbital solutions. Of the rest, nine were new to us, but reports of variability detection from ground-based surveys exists for seven of these, while two are completely new. Although these new systems are certainly worth closer follow-up, none of them are in the continuous viewing zone of \tess, and therefore the span of the observations is limited to one or two months. We note that the systems with periods comparable to or longer than that of \aador\ would be particularly welcome targets for the \romer\ delay measurements, since their light-travel-time delays would be more significant in the wider orbits, and therefore easier to measure more accurately. We show the folded \tess\ light curves of the nine new HW-Vir systems in Figure\,\ref{fig:hwvirs}.

\subsection{J19447+5449}
TIC\,467187065 (GALEX\,J194442.8+544942), with a period of 0.0642\,d is among the shortest-period eclipsing sdB+dM systems known. The system is listed as an HW-Vir candidate in the EREBOS project \citep{schaffenroth19}, but no details are provided. Unfortunately, we have no spectroscopic data yet for this star, but the \tess\ light-curve can leave little doubt that the system must be a sdB+dM binary. The folded light curve shown in the first panel of Figure\,\ref{fig:hwvirs} shows that the eclipses are fairly deep and appear flat-bottomed, while the secondary eclipses are completely obscuring the reflected light from the companion, indicating that the subdwarf is large enough to fully eclipse the companion for about 10\% of the orbital period. Thus, the companion might be a brown dwarf. The Gaia EDR3 parallax of 0.664(36)\,mas is as would be expected for a sdB star with a G-band magnitude of 15.75, leaving little room for doubt about its nature.

\begin{figure}
\includegraphics[width=\hsize]{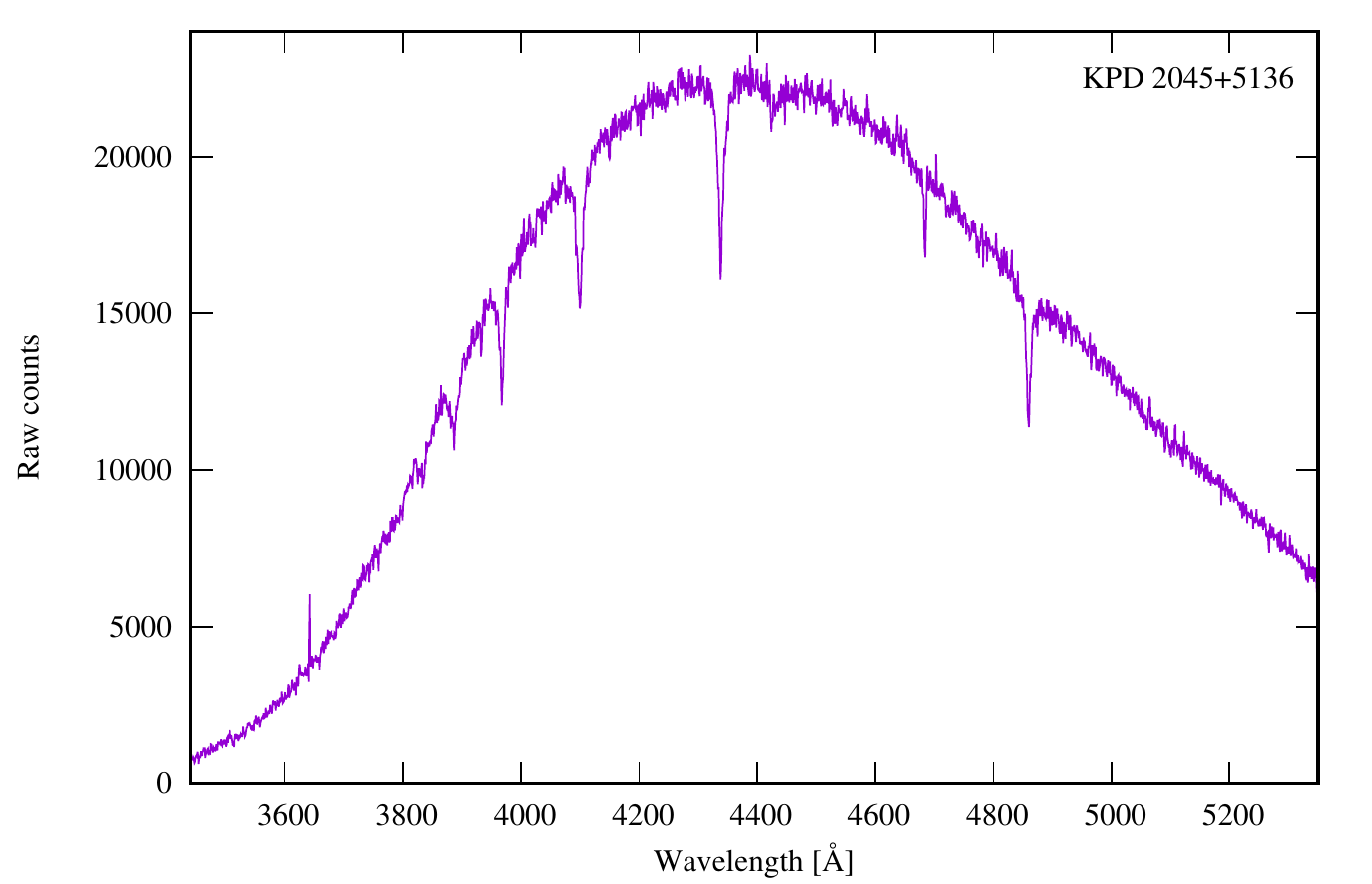}
\caption{The spectrum of KPD\,2045+5136 clearly identifies the primary as a sdO star.}
\label{fig:kpd2045}
\end{figure}

\subsection{KPD\,2045+5136}
TIC\,365213081, with a period of 0.0896\,d is another very short-period sdB+dM binary. It was identified as a sdB star in the Kitt-Peak Downes survey \citep{downes86}, but no further details have been published. We obtained a spectrum of the target with the 2.56-m Nordic Optical Telescope on 16 Jan 2020 using the ALFOSC spectrograph. The spectrum is shown in Fig.\,\ref{fig:kpd2045}, and clearly presents the strong He-II line at 4686\,\AA and no He-I line at 4471\,\AA, which identifies this as a sdO star. The Gaia parallax of 0.931(23)\,mas is somewhat high for a G\,=\,15.23 mag sdO, but not unreasonable considering that the object is only 5$^\circ$ away from the Galactic plane and therefore has a rather high reddening, E(B-V)\,=\,1.07 \citep{geier17}. KPD\,2045+5136 was observed by \tess\ in Sectors 15 and 16. The light curve (second panel in Figure\,\ref{fig:hwvirs}) shows a very shallow primary eclipse, and marginal secondary eclipse, indicating a rather low inclination angle causing grazing eclipses.

\subsection{HE\,0516--2311}
TIC\,408187719, with P\,=\,0.0912\,d is also at the very-short end of the sdB+dM period distribution. It was included in the kinematic survey of \citet{altmann04}, and also occurs in the EC survey Zone 2 as EC\,05160--2311 \citep{odonoghue13}. \citet{2003PhDT........48E} used low resolution (5.3\AA) spectra covering a useful spectral range of 3500 -- 5500\AA\ obtained with the DFOSC spectrograph at 1.54-m Danish telescope (ESO, La Silla) and derived \teff = 30\,100$\pm$300\,K, \logg = 5.5$\pm$0.1 and a helium-to-hydrogen ratio y of $\log$ y = -2.0$\pm$0.1, typical for sdB stars. The target was observed with \tess\ in 2-minute mode in Sectors 5 and 6. The light curve (third panel of Figure\,\ref{fig:hwvirs}) shows a typical HW-Vir-type shape, but the eclipse depths are shallow. 

\subsection{J19065+2807}
TIC\,281948821 (RA\,=\,19:06:35.6, Dec\,=\,+28:07:20.8) is a new system. Its Gaia EDR3 parallax is 0.637(30)\,mas, G\,=\,15.64, and it was observed in Sector 14. No spectroscopy is available. It is included in the EREBOS candidate list as J286.6485+28.1219. The period of P\,=\,0.1121\,d is just 4\% shorter than that of the class prototype. The V-shaped eclipses are exceptionally deep, reaching a flux of only 15\% of the mean level during a primary eclipse, which indicates that the two companions have almost the same radius, and that the inclination angle is close to 90$^\circ$.

\subsection{EVR-CB-003}
\begin{figure*}
\includegraphics[width=\hsize]{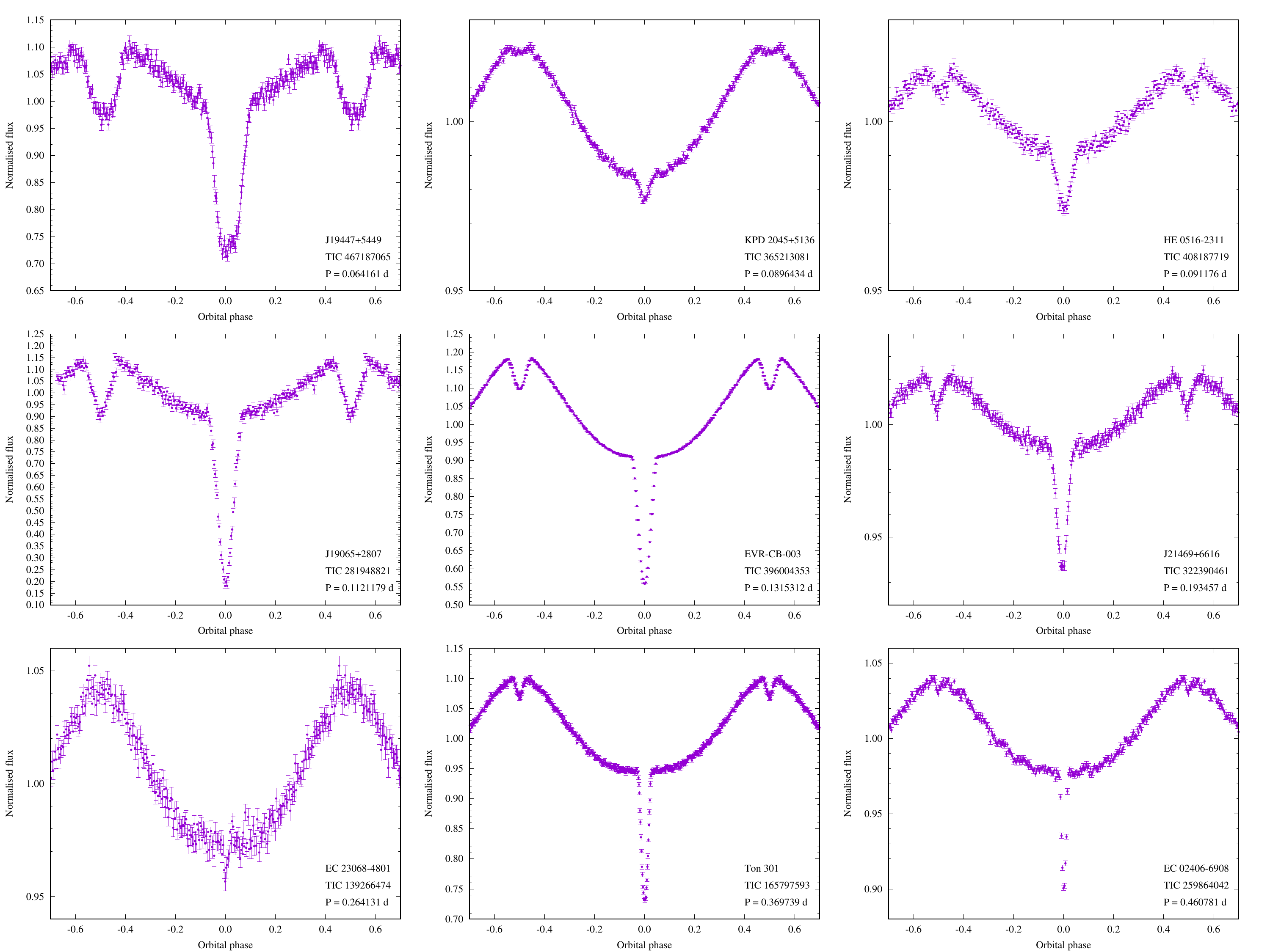}
\caption{The folded light curves of nine HW\,Vir candidate systems we found in \tess\ photometry.}
\label{fig:hwvirs}
\end{figure*}

TIC\,396004353 (GALEX\,J140155.3-751333, J14019--7513) was identified as a hot subdwarf star from a survey of GALEX selected UV-bright objects included in \citet{geier17}. The low-resolution NTT/EFOSC survey spectrum taken in June 2015 confirms that the primary is a sdB star, with T$_{\rm eff}$\,$\sim$\,30\,kK, $\logg$\,=\,5.6 and no detectable helium. It is reported as a new HW-Vir system dubbed EVR-CB-003 in the recent Evryscope survey \citep{ratzloff20}, with a promise of a forthcoming discovery paper. It was observed with \tess\ in Sectors 11 and 12, and owing to the brightness of this object (G\,=\,13.5), the light curve is of excellent precision (middle panel in Figure\,\ref{fig:hwvirs}). With a period (P\,=\,0.1315\,d), only slightly longer than that of the class prototype, the light curve appears almost identical in shape.

\subsection{J21469+6616}
TIC\,322390461 (RA\,=\,21:46:56.6, Dec\,=\,66:16:06.9) is a new system. With a period of 0.1935\,d it is in the middle of the range of known HW\,Vir systems. The object is rather faint at G\,=\,16.22, and with a parallax of 0.680(35)\,mas and B-R color of 0.37\,mag, the object must be highly reddened. Thus, the \tess\ light curve of TIC\,322390461, observed in 2-minute cadence in Sectors 16 and 18 is rather noisy, and the eclipse depths likely underestimated by about 50\%\ due to contamination. Ground-based follow-up would be wanted in order to explore the details of this system.

\subsection{EC\,23068$-$4801}
TIC\,139266474 was observed by \tess\ in Sector\,1, and has a rather long period of P\,=\,0.2641, which is just 1\% longer than that of \aador. It is listed as a rotationally variable object in the Catalina survey \citep{drake17}, but not specifically mentioned. The \tess\ light curve of TIC\,139266474 shows the shallowest gracing primary eclipse seen so far, while the secondary eclipse is not detectable, making this a poor target for time-delay measurements.

\subsection{Ton\,301}
TIC\,165797593 was first classified as an RR\,Lyr variable in the Catalina survey, and correctly identified as a \hwvir\ candidate by the EREBOS team. The period is one of the longest found for sdB+dM systems, at P\,=\,0.3697\,d. At G\,=\,13.80, the light curve is of good quality, and quite similar to that of \aador. A Fourier amplitude spectrum shows a few peaks in the g-mode region of sdB stars, indicating that the primary is likely a pulsator. Ton\,301 is an excellent new target for monitoring by the Northern observatories.

\subsection{EC\,02406$-$6908}
TIC\,259864042 was observed by \tess\ in Sectors 1 and 2. With P\,=\,0.4607\,d it is the longest period HW-Vir system detected to date. With the period almost twice as long as \aador\ the \romer\ delay should be correspondingly longer. However, eclipses are rather shallow, indicating a lower inclination angle than that of \aador, and this will make timing measurements less accurate. Furthermore, the folded light curve appears more noisy than expected for its magnitude (G\,=\,14.67), and inspecting a Fourier amplitude spectrum reveals several peaks interleaved between the orbital harmonics, which implies that the primary is a g-mode pulsator like \hwvir. This target is definitely deserving of a follow-up study.

\section{Summary}
We have presented new photometric observations of \aador\ collected with the \tess\ satellite. Since \aador\ is located in the southern continuous viewing zone of \tess, we obtained a total of 12 sectors of data spanning just over one year. \aador\ is a close binary system consisting of a hot primary and a cool companion. The light curve shows variations caused by mutual eclipsing and an irradiation effect. As in other \hwvirs, we searched for stellar pulsations characteristic of the hot primary. We detected just one peak with FAP\,=\,0.2\%, which makes \aador\ primary a promising candidate for sdBV. This detection should be confirmed with 20\,sec cadences currently being collected. The amplitude of this potential pulsation peak is barely exceeding our detection threshold. The reason may be twofold. The primary is hotter than those in other systems, which does not favor the driving mechanism, and the \tess\ data precision is lower than \kep\ data, resulting in a higher detection limit.

We have analysed eclipses by deriving their mid-times and calculating an O--C diagram. This analysis confirmed a very stable orbital period with a limit on the period change to be not faster than 5.75$\cdot$10$^{-13}$\,s/s. Such a value excludes a tertiary body in the system, as well as any significant mass exchange. We have also measured a shift of the secondary eclipse. There are only a few other sdB stars for which the shift of the secondary eclipse has been derived. \cite{baran15a} reported a shift of 1.76\,s for \toby\ after it had been observed for three years with the \kep\ spacecraft, \cite{baran18} reported a shift of 1.509\,sec for \hwvir\ based on {\it K2} data, and another two stars were reported by \cite{schaffenroth15} and \cite{lee17} using ground-based data, yet neither had sufficient precision to derive a precise shift. The shifts in \toby\ and \hwvir\, under an assumption of circular orbits, are too small to reproduce canonical masses for the primaries. It was suggested that either the mass is truly low and the primaries are low-mass post RGB stars, which have not gone through a helium flash, or the shifts are affected by non-zero eccentricity that is too small to be confirmed by current observations. \aador\ is the first case where the shift agrees with the prediction based on a radial velocity amplitude of the cool companion. Such a radial velocity amplitude has not yet been derived either for \toby\ or for \hwvir. We note that the wider orbit of \aador\ compared to the other systems is favorable for reliable \romer-delay measurements, not just because the delay itself is longer and therefore easier to measure, but because the higher-order effects caused by the large size of the stellar bodies compared to the size of the orbit (not accounted by \cite{kaplan10}) become smaller, as discussed by \cite{baran18}.

The fundamental stellar parameters (radius, mass, and luminosity) were derived independently through analysis of the spectral energy distribution by making use of the high-precision trigonometric parallax from Gaia. The results of both approaches corroborate that the primary of AA\,Dor is a canonical mass hot subdwarf, and its effective temperature, radius and luminosity are consistent with an evolutionary state beyond core helium burning. The mass of the secondary companion is close to the limit for nuclear burning.

We searched the public TESS archive for HW-Vir candidates by matching targets with the  \gaia-selected subdwarf sample \citep{geier17}. Although these stars are certainly worth closer follow-up, none of them are in the continuous viewing zone, and therefore the length of the \tess\ observations is limited to one or two months. However, \tess\ will revisit these systems in its extended mission, hopefully several times. In combination with ground-based follow-up with high-precision photometry, it can be possible to apply the \romer\ delay method on several of these systems, especially those systems that have periods longer than \aador.

\section*{Acknowledgements}
The authors thank Dave Kilkenny for providing the newest unpublished orbital period estimation included in Table\,\ref{tab:pastperiod}, as well as Kosmas Gazeas and Brad Barlow for generous comments. Financial support from the Polish National Science Centre under projects No.\,UMO-2017/26/E/ST9/00703 and UMO-2017/25/B/ST9/02218 is appreciated. A.I.\ and U.H.\ acknowledge funding by the Deutsche For\-schungs\-gemeinschaft (DFG) through grants IR190/1-1, HE1356/70-1, and HE1356/71-1. The spectroscopic observations used in this work were obtained with the Nordic Optical Telescope at the Observatorio del Roque de los Muchachos and operated jointly by Denmark, Finland, Iceland, Norway, and Sweden. This work has made use of data from the European Space Agency (ESA) mission {\it Gaia} (\url{https://www.cosmos.esa.int/gaia}), processed by the {\it Gaia} Data Processing and Analysis Consortium (DPAC, \url{https://www.cosmos.esa.int/web/gaia/dpac/consortium}). Funding for the DPAC has been provided by national institutions, in particular the institutions participating in the {\it Gaia} Multilateral Agreement. This research was made possible through the use of the AAVSO Photometric All-Sky Survey (APASS), funded by the Robert Martin Ayers Sciences Fund. This paper includes data collected by the \tess\ mission. Funding for the \tess\ mission is provided by the NASA Explorer Program.

\section*{Data availability}
The datasets were derived from MAST in the public domain archive.stsci.edu and ESA mission {\it Gaia} accessible at https://www.cosmos.esa.int/gaia.

\bibliography{myrefs}

\begin{thebibliography}{}
\makeatletter
\relax
\def\mn@urlcharsother{\let\do\@makeother \do\$\do\&\do\#\do\^\do\_\do\%\do\~}
\def\mn@doi{\begingroup\mn@urlcharsother \@ifnextchar [ {\mn@doi@}
  {\mn@doi@[]}}
\def\mn@doi@[#1]#2{\def\@tempa{#1}\ifx\@tempa\@empty \href
  {http://dx.doi.org/#2} {doi:#2}\else \href {http://dx.doi.org/#2} {#1}\fi
  \endgroup}
\def\mn@eprint#1#2{\mn@eprint@#1:#2::\@nil}
\def\mn@eprint@arXiv#1{\href {http://arxiv.org/abs/#1} {{\tt arXiv:#1}}}
\def\mn@eprint@dblp#1{\href {http://dblp.uni-trier.de/rec/bibtex/#1.xml}
  {dblp:#1}}
\def\mn@eprint@#1:#2:#3:#4\@nil{\def\@tempa {#1}\def\@tempb {#2}\def\@tempc
  {#3}\ifx \@tempc \@empty \let \@tempc \@tempb \let \@tempb \@tempa \fi \ifx
  \@tempb \@empty \def\@tempb {arXiv}\fi \@ifundefined
  {mn@eprint@\@tempb}{\@tempb:\@tempc}{\expandafter \expandafter \csname
  mn@eprint@\@tempb\endcsname \expandafter{\@tempc}}}

\bibitem[\protect\citeauthoryear{{Almeida}, {Jablonski}, {Tello}  \&
  {Rodrigues}}{{Almeida} et~al.}{2012}]{almeida12}
{Almeida} L.~A.,  {Jablonski} F.,  {Tello} J.,   {Rodrigues} C.~V.,  2012,
  \mn@doi [\mnras] {10.1111/j.1365-2966.2012.20891.x}, \href
  {https://ui.adsabs.harvard.edu/abs/2012MNRAS.423..478A} {423, 478}

\bibitem[\protect\citeauthoryear{{Altmann}, {Edelmann}  \& {de Boer}}{{Altmann}
  et~al.}{2004}]{altmann04}
{Altmann} M.,  {Edelmann} H.,   {de Boer} K.~S.,  2004, \mn@doi [\aap]
  {10.1051/0004-6361:20031606}, \href
  {https://ui.adsabs.harvard.edu/abs/2004A&A...414..181A} {414, 181}

\bibitem[\protect\citeauthoryear{Baran, Zola, Blokesz, {\O}stensen  \&
  Silvotti}{Baran et~al.}{2015}]{baran15a}
Baran A.,  Zola S.,  Blokesz A.,  {\O}stensen R.,   Silvotti R.,  2015, A\&A,
  577, 146

\bibitem[\protect\citeauthoryear{Baran et~al.,}{Baran et~al.}{2018}]{baran18}
Baran A.,  et~al., 2018, MNRAS, 481, 2721

\bibitem[\protect\citeauthoryear{Baran, Telting, Jeffery, {\O}stensen, Vos,
  Reed  \& Vu{\v c}kovi{\'c}}{Baran et~al.}{2019}]{baran19}
Baran A.,  Telting J.,  Jeffery C.,  {\O}stensen R.,  Vos J.,  Reed M.,   Vu{\v
  c}kovi{\'c} M.,  2019, MNRAS, 489, 1556

\bibitem[\protect\citeauthoryear{Barlow, Wade  \& Liss}{Barlow
  et~al.}{2012}]{barlow12}
Barlow B.,  Wade R.,   Liss S.,  2012, ApJ, 753, 101

\bibitem[\protect\citeauthoryear{{Barlow} et~al.,}{{Barlow}
  et~al.}{2013}]{barlow13}
{Barlow} B.~N.,  et~al., 2013, \mn@doi [\mnras] {10.1093/mnras/sts271}, \href
  {http://cdsads.u-strasbg.fr/abs/2013MNRAS.430...22B} {430, 22}

\bibitem[\protect\citeauthoryear{Brown, Landsman, Randall, Sweigart  \&
  T.}{Brown et~al.}{2013}]{brown13}
Brown T.,  Landsman W.,  Randall S.,  Sweigart A.,   T. L.,  2013, ApJ, 777, 22

\bibitem[\protect\citeauthoryear{Charpinet, Fontaine, Brassard, Chayer, Rogers,
  Iglesias  \& Dorman}{Charpinet et~al.}{1997}]{charpinet97}
Charpinet S.,  Fontaine G.,  Brassard P.,  Chayer P.,  Rogers F.~J.,  Iglesias
  C.~A.,   Dorman B.,  1997, ApJ, 483, 123

\bibitem[\protect\citeauthoryear{Charpinet et~al.,}{Charpinet
  et~al.}{2011}]{charpinet11}
Charpinet S.,  et~al., 2011, A\&A, 530, 3

\bibitem[\protect\citeauthoryear{Charpinet et~al.,}{Charpinet
  et~al.}{2019}]{charpinet19}
Charpinet S.,  et~al., 2019, \mn@doi [A\&A] {10.1051/0004-6361/201935395}, 632,
  90

\bibitem[\protect\citeauthoryear{{Cutri} et~al.,}{{Cutri}
  et~al.}{2003}]{2003yCat.2246....0C}
{Cutri} R.~M.,  et~al., 2003, VizieR Online Data Catalog, \href
  {https://ui.adsabs.harvard.edu/abs/2003yCat.2246....0C} {p. II/246}

\bibitem[\protect\citeauthoryear{{Dorman}, {Rood}  \& {O'Connell}}{{Dorman}
  et~al.}{1993}]{dorman93}
{Dorman} B.,  {Rood} R.~T.,   {O'Connell} R.~W.,  1993, \mn@doi [\apj]
  {10.1086/173511}, \href {http://adsabs.harvard.edu/abs/1993ApJ...419..596D}
  {419, 596}

\bibitem[\protect\citeauthoryear{{Downes}}{{Downes}}{1986}]{downes86}
{Downes} R.~A.,  1986, \mn@doi [\apjs] {10.1086/191124}, \href
  {http://cdsads.u-strasbg.fr/abs/1986ApJS...61..569D} {61, 569}

\bibitem[\protect\citeauthoryear{{Drake} et~al.,}{{Drake}
  et~al.}{2017}]{drake17}
{Drake} A.~J.,  et~al., 2017, \mn@doi [\mnras] {10.1093/mnras/stx1085}, \href
  {https://ui.adsabs.harvard.edu/abs/2017MNRAS.469.3688D} {469, 3688}

\bibitem[\protect\citeauthoryear{{Drechsel} et~al.,}{{Drechsel}
  et~al.}{2001}]{drechsel01}
{Drechsel} H.,  et~al., 2001, \mn@doi [\aap] {10.1051/0004-6361:20011376},
  \href {http://cdsads.u-strasbg.fr/abs/2001A%26A...379..893D} {379, 893}

\bibitem[\protect\citeauthoryear{{Edelmann}}{{Edelmann}}{2003}]{2003PhDT........48E}
{Edelmann} H.,  2003, PhD thesis, Friedrich-Alexander University
  Erlangen-N{\"u}rnberg
  \url{https://www.sternwarte.uni-erlangen.de/docs/theses/2003-07_Edelmann.pdf}

\bibitem[\protect\citeauthoryear{{Fitzpatrick}, {Massa}, {Gordon}, {Bohlin}  \&
  {Clayton}}{{Fitzpatrick} et~al.}{2019}]{2019ApJ...886..108F}
{Fitzpatrick} E.~L.,  {Massa} D.,  {Gordon} K.~D.,  {Bohlin} R.,   {Clayton}
  G.~C.,  2019, \mn@doi [\apj] {10.3847/1538-4357/ab4c3a}, \href
  {https://ui.adsabs.harvard.edu/abs/2019ApJ...886..108F} {886, 108}

\bibitem[\protect\citeauthoryear{{Fleig}, {Rauch}, {Werner}  \& {Kruk}}{{Fleig}
  et~al.}{2008}]{2008A&A...492..565F}
{Fleig} J.,  {Rauch} T.,  {Werner} K.,   {Kruk} J.~W.,  2008, \mn@doi [\aap]
  {10.1051/0004-6361:200810738}, \href
  {https://ui.adsabs.harvard.edu/abs/2008A&A...492..565F} {492, 565}

\bibitem[\protect\citeauthoryear{Fontaine, Brassard, Charpinet, Green, Randall
  \& Van~Grootel}{Fontaine et~al.}{2012}]{fontaine12}
Fontaine G.,  Brassard P.,  Charpinet S.,  Green E.~M.,  Randall S.~K.,
  Van~Grootel V.,  2012, A\&A, 539, 12

\bibitem[\protect\citeauthoryear{{For} et~al.,}{{For} et~al.}{2010}]{for10}
{For} B.-Q.,  et~al., 2010, \mn@doi [\apj] {10.1088/0004-637X/708/1/253}, \href
  {http://cdsads.u-strasbg.fr/abs/2010ApJ...708..253F} {708, 253}

\bibitem[\protect\citeauthoryear{{Fouqu{\'e}} et~al.,}{{Fouqu{\'e}}
  et~al.}{2000}]{2000A&AS..141..313F}
{Fouqu{\'e}} P.,  et~al., 2000, \mn@doi [\aaps] {10.1051/aas:2000123}, \href
  {https://ui.adsabs.harvard.edu/abs/2000A&AS..141..313F} {141, 313}

\bibitem[\protect\citeauthoryear{{Gaia Collaboration} et~al.,}{{Gaia
  Collaboration} et~al.}{2016}]{2016A&A...595A...1G}
{Gaia Collaboration} et~al., 2016, \mn@doi [\aap]
  {10.1051/0004-6361/201629272}, \href
  {https://ui.adsabs.harvard.edu/abs/2016A&A...595A...1G} {595, A1}

\bibitem[\protect\citeauthoryear{{Gaia Collaboration}, {Brown}, {Vallenari},
  {Prusti}, {de Bruijne}, {Babusiaux}  \& {Biermann}}{{Gaia Collaboration}
  et~al.}{2020}]{2020arXiv201201533G}
{Gaia Collaboration} {Brown} A.~G.~A.,  {Vallenari} A.,  {Prusti} T.,  {de
  Bruijne} J.~H.~J.,  {Babusiaux} C.,   {Biermann} M.,  2020, arXiv e-prints,
  \href {https://ui.adsabs.harvard.edu/abs/2020arXiv201201533G} {p.
  arXiv:2012.01533}

\bibitem[\protect\citeauthoryear{{Geier} et~al.,}{{Geier}
  et~al.}{2011}]{geier11}
{Geier} S.,  et~al., 2011, \mn@doi [\apjl] {10.1088/2041-8205/731/2/L22}, \href
  {http://cdsads.u-strasbg.fr/abs/2011ApJ...731L..22G} {731, L22}

\bibitem[\protect\citeauthoryear{{Geier}, {{\O}stensen}, {Nemeth}, {Gentile
  Fusillo}, {G{\"a}nsicke}, {Telting}, {Green}  \& {Schaffenroth}}{{Geier}
  et~al.}{2017}]{geier17}
{Geier} S.,  {{\O}stensen} R.~H.,  {Nemeth} P.,  {Gentile Fusillo} N.~P.,
  {G{\"a}nsicke} B.~T.,  {Telting} J.~H.,  {Green} E.~M.,   {Schaffenroth} J.,
  2017, \mn@doi [\aap] {10.1051/0004-6361/201630135}, \href
  {https://ui.adsabs.harvard.edu/abs/2017A&A...600A..50G} {600, A50}

\bibitem[\protect\citeauthoryear{Han, Podsiadlowski, Maxted, Marsh  \&
  Ivanova}{Han et~al.}{2002}]{han02}
Han Z.,  Podsiadlowski P.,  Maxted P. F.~L.,  Marsh T.~R.,   Ivanova N.,  2002,
  MNRAS, 336, 449

\bibitem[\protect\citeauthoryear{Han, Podsiadlowski, Maxted  \& Marsh}{Han
  et~al.}{2003}]{han03}
Han Z.,  Podsiadlowski P.,  Maxted P. F.~L.,   Marsh T.~R.,  2003, MNRAS, 341,
  669

\bibitem[\protect\citeauthoryear{{Hansen} \& {Spangenberg}}{{Hansen} \&
  {Spangenberg}}{1971}]{1971ApJ...163..653H}
{Hansen} C.~J.,  {Spangenberg} W.,  1971, \mn@doi [\apj] {10.1086/150808},
  \href {https://ui.adsabs.harvard.edu/abs/1971ApJ...163..653H} {163, 653}

\bibitem[\protect\citeauthoryear{{Hauck} \& {Mermilliod}}{{Hauck} \&
  {Mermilliod}}{1998}]{1998A&AS..129..431H}
{Hauck} B.,  {Mermilliod} M.,  1998, \mn@doi [\aaps] {10.1051/aas:1998195},
  \href {https://ui.adsabs.harvard.edu/abs/1998A&AS..129..431H} {129, 431}

\bibitem[\protect\citeauthoryear{Heber}{Heber}{2016}]{heber16}
Heber U.,  2016, PASP, 128, 2001

\bibitem[\protect\citeauthoryear{{Heber}}{{Heber}}{2017}]{2017ASPC..509...85H}
{Heber} U.,  2017, in {Tremblay} P.~E.,  {Gaensicke} B.,   {Marsh} T.,  eds,
  Astronomical Society of the Pacific Conference Series Vol. 509, 20th European
  White Dwarf Workshop. p.~85 (\mn@eprint {arXiv} {1610.07516})

\bibitem[\protect\citeauthoryear{{Heber}, {Irrgang}  \& {Schaffenroth}}{{Heber}
  et~al.}{2018}]{2018OAst...27...35H}
{Heber} U.,  {Irrgang} A.,   {Schaffenroth} J.,  2018, \mn@doi [Open Astronomy]
  {10.1515/astro-2018-0008}, \href
  {https://ui.adsabs.harvard.edu/abs/2018OAst...27...35H} {27, 35}

\bibitem[\protect\citeauthoryear{{Henden}, {Levine}, {Terrell}  \&
  {Welch}}{{Henden} et~al.}{2015}]{2015AAS...22533616H}
{Henden} A.~A.,  {Levine} S.,  {Terrell} D.,   {Welch} D.~L.,  2015, in
  American Astronomical Society Meeting Abstracts \#225. p. 336.16

\bibitem[\protect\citeauthoryear{{Hilditch}, {Harries}  \& {Hill}}{{Hilditch}
  et~al.}{1996}]{hilditch96}
{Hilditch} R.~W.,  {Harries} T.~J.,   {Hill} G.,  1996, \mn@doi [\mnras]
  {10.1093/mnras/279.4.1380}, \href
  {http://adsabs.harvard.edu/abs/1996MNRAS.279.1380H} {279, 1380}

\bibitem[\protect\citeauthoryear{{Hilditch}, {Kilkenny}, {Lynas-Gray}  \&
  {Hill}}{{Hilditch} et~al.}{2003}]{2003MNRAS.344..644H}
{Hilditch} R.~W.,  {Kilkenny} D.,  {Lynas-Gray} A.~E.,   {Hill} G.,  2003,
  \mn@doi [\mnras] {10.1046/j.1365-8711.2003.06860.x}, \href
  {https://ui.adsabs.harvard.edu/abs/2003MNRAS.344..644H} {344, 644}

\bibitem[\protect\citeauthoryear{{H{\o}g} et~al.,}{{H{\o}g}
  et~al.}{2000}]{2000A&A...355L..27H}
{H{\o}g} E.,  et~al., 2000, \aap, \href
  {https://ui.adsabs.harvard.edu/abs/2000A&A...355L..27H} {355, L27}

\bibitem[\protect\citeauthoryear{Hoyer, Rauch, Werner, Hauschildt  \&
  Kruk}{Hoyer et~al.}{2015}]{hoyer15}
Hoyer D.,  Rauch T.,  Werner K.,  Hauschildt P.,   Kruk J.,  2015, A\&A, 578,
  125

\bibitem[\protect\citeauthoryear{{Hu}, {Dupret}, {Aerts}, {Nelemans},
  {Kawaler}, {Miglio}, {Montalban}  \& {Scuflaire}}{{Hu}
  et~al.}{2008}]{2008A&A...490..243H}
{Hu} H.,  {Dupret} M.~A.,  {Aerts} C.,  {Nelemans} G.,  {Kawaler} S.~D.,
  {Miglio} A.,  {Montalban} J.,   {Scuflaire} R.,  2008, \mn@doi [\aap]
  {10.1051/0004-6361:200810233}, \href
  {https://ui.adsabs.harvard.edu/abs/2008A&A...490..243H} {490, 243}

\bibitem[\protect\citeauthoryear{Kaplan}{Kaplan}{2010}]{kaplan10}
Kaplan D.,  2010, ApJ, 717, 108

\bibitem[\protect\citeauthoryear{Kilkenny}{Kilkenny}{1983}]{kilkenny83}
Kilkenny D.,  1983, SAAOC, 7, 55

\bibitem[\protect\citeauthoryear{Kilkenny}{Kilkenny}{1986}]{kilkenny86}
Kilkenny D.,  1986, The Observatory, 106, 160

\bibitem[\protect\citeauthoryear{Kilkenny}{Kilkenny}{2011}]{kilkenny11}
Kilkenny D.,  2011, MNRAS, 412, 487

\bibitem[\protect\citeauthoryear{Kilkenny \& Hill}{Kilkenny \&
  Hill}{1975}]{kilkenny75}
Kilkenny D.,  Hill P.~W.,  1975, \mn@doi [MNRAS] {10.1093/mnras/173.3.625},
  173, 625

\bibitem[\protect\citeauthoryear{Kilkenny, Hilditch  \& Penfold}{Kilkenny
  et~al.}{1978}]{kilkenny78}
Kilkenny D.,  Hilditch R.,   Penfold J.,  1978, MNRAS, 183, 523

\bibitem[\protect\citeauthoryear{Kilkenny, Lynas-Gray  \& Hilditch}{Kilkenny
  et~al.}{1979}]{kilkenny79}
Kilkenny D.,  Lynas-Gray A.,   Hilditch R.,  1979, in van Horn H.,  Weidemann
  V.,  eds, White Dwarfs and Variable Degenerate Stars. Proceedings of IAU
  Colloq. 53.
p.~255

\bibitem[\protect\citeauthoryear{Kilkenny, Harrop-Allin  \& Marang}{Kilkenny
  et~al.}{1991}]{kilkenny91}
Kilkenny D.,  Harrop-Allin M.,   Marang F.,  1991, IBVS, 3569, 1

\bibitem[\protect\citeauthoryear{Kilkenny, Koen, O'Donoghue  \&
  Stobie}{Kilkenny et~al.}{1997}]{kilkenny97}
Kilkenny D.,  Koen C.,  O'Donoghue D.,   Stobie R.~S.,  1997, MNRAS, 285, 640

\bibitem[\protect\citeauthoryear{Kilkenny, Keuris, Marang, Roberts, van Wyk  \&
  Ogloza}{Kilkenny et~al.}{2000}]{kilkenny00}
Kilkenny D.,  Keuris S.,  Marang F.,  Roberts G.,  van Wyk F.,   Ogloza W.,
  2000, The Observatory, 120, 48

\bibitem[\protect\citeauthoryear{{Klepp} \& {Rauch}}{{Klepp} \&
  {Rauch}}{2011}]{2011A&A...531L...7K}
{Klepp} S.,  {Rauch} T.,  2011, \mn@doi [\aap] {10.1051/0004-6361/201116887},
  \href {https://ui.adsabs.harvard.edu/abs/2011A&A...531L...7K} {531, L7}

\bibitem[\protect\citeauthoryear{Kwee \& van Woerden}{Kwee \& van
  Woerden}{1956}]{kwee56}
Kwee K.,  van Woerden H.,  1956, Bulletin of the Astronomical Institutes of the
  Netherlands, 12, 327

\bibitem[\protect\citeauthoryear{Lee, Youn, Hong  \& Han}{Lee
  et~al.}{2017}]{lee17}
Lee J.,  Youn J.-H.,  Hong K.,   Han W.,  2017, ApJ, 839, 39

\bibitem[\protect\citeauthoryear{{Lindegren} et~al.,}{{Lindegren}
  et~al.}{2020}]{2020arXiv201203380L}
{Lindegren} L.,  et~al., 2020, arXiv e-prints, \href
  {https://ui.adsabs.harvard.edu/abs/2020arXiv201203380L} {p. arXiv:2012.03380}

\bibitem[\protect\citeauthoryear{Menzies \& Marang}{Menzies \&
  Marang}{1986}]{menzies86}
Menzies J.,  Marang F.,  1986, in Hearnshaw J.,  Cottrell P.,  Reidel
  Dordrecht eds, Instrumentation and Research Programmes for Small Telescopes.
  Proceedings of the International Astronomical Union Symposium No. 118.
p.~305

\bibitem[\protect\citeauthoryear{{O'Donoghue}, {Kilkenny}, {Koen}, {Hambly},
  {MacGillivray}  \& {Stobie}}{{O'Donoghue} et~al.}{2013}]{odonoghue13}
{O'Donoghue} D.,  {Kilkenny} D.,  {Koen} C.,  {Hambly} N.,  {MacGillivray} H.,
   {Stobie} R.~S.,  2013, \mn@doi [\mnras] {10.1093/mnras/stt158}, \href
  {https://ui.adsabs.harvard.edu/abs/2013MNRAS.431..240O} {431, 240}

\bibitem[\protect\citeauthoryear{{Onken} et~al.,}{{Onken}
  et~al.}{2019}]{2019PASA...36...33O}
{Onken} C.~A.,  et~al., 2019, \mn@doi [\pasa] {10.1017/pasa.2019.27}, \href
  {https://ui.adsabs.harvard.edu/abs/2019PASA...36...33O} {36, e033}

\bibitem[\protect\citeauthoryear{{\O}stensen et~al.,}{{\O}stensen
  et~al.}{2010}]{ostensen10}
{\O}stensen R.,  et~al., 2010, MNRAS, 408, 51

\bibitem[\protect\citeauthoryear{{Ostrowski}, {Baran}, {Sanjayan}  \&
  {Sahoo}}{{Ostrowski} et~al.}{2021}]{2020arXiv201114621O}
{Ostrowski} J.,  {Baran} A.,  {Sanjayan} S.,   {Sahoo} S.~K.,  2021, arXiv
  e-prints, \href {https://ui.adsabs.harvard.edu/abs/2020arXiv201114621O} {p.
  arXiv:2011.14621}

\bibitem[\protect\citeauthoryear{{Pribulla} et~al.,}{{Pribulla}
  et~al.}{2013}]{pribulla13}
{Pribulla} T.,  et~al., 2013, Information Bulletin on Variable Stars, \href
  {https://ui.adsabs.harvard.edu/abs/2013IBVS.6067....1P} {6067, 1}

\bibitem[\protect\citeauthoryear{{Ratzloff} et~al.,}{{Ratzloff}
  et~al.}{2020}]{ratzloff20}
{Ratzloff} J.~K.,  et~al., 2020, \mn@doi [\apj] {10.3847/1538-4357/ab64f3},
  \href {https://ui.adsabs.harvard.edu/abs/2020ApJ...890..126R} {890, 126}

\bibitem[\protect\citeauthoryear{{Rauch}}{{Rauch}}{2000}]{2000A&A...356..665R}
{Rauch} T.,  2000, \aap, \href
  {https://ui.adsabs.harvard.edu/abs/2000A&A...356..665R} {356, 665}

\bibitem[\protect\citeauthoryear{Rauch \& Werner}{Rauch \&
  Werner}{2003}]{rauch03}
Rauch T.,  Werner K.,  2003, A\&A, 400, 271

\bibitem[\protect\citeauthoryear{{Reed}}{{Reed}}{2003}]{2003AJ....125.2531R}
{Reed} B.~C.,  2003, \mn@doi [\aj] {10.1086/374771}, \href
  {https://ui.adsabs.harvard.edu/abs/2003AJ....125.2531R} {125, 2531}

\bibitem[\protect\citeauthoryear{{Riello} et~al.,}{{Riello}
  et~al.}{2020}]{2020arXiv201201916R}
{Riello} M.,  et~al., 2020, arXiv e-prints, \href
  {https://ui.adsabs.harvard.edu/abs/2020arXiv201201916R} {p. arXiv:2012.01916}

\bibitem[\protect\citeauthoryear{{Schaffenroth}, {Geier}, {Drechsel}, {Heber},
  {Wils}, {{\O}stensen}, {Maxted}  \& {di Scala}}{{Schaffenroth}
  et~al.}{2013}]{schaffenroth13}
{Schaffenroth} V.,  {Geier} S.,  {Drechsel} H.,  {Heber} U.,  {Wils} P.,
  {{\O}stensen} R.~H.,  {Maxted} P.~F.~L.,   {di Scala} G.,  2013, \mn@doi
  [\aap] {10.1051/0004-6361/201220929}, \href
  {https://ui.adsabs.harvard.edu/abs/2013A&A...553A..18S} {553, A18}

\bibitem[\protect\citeauthoryear{{Schaffenroth}, {Geier}, {Heber}, {Kupfer},
  {Ziegerer}, {Heuser}, {Classen}  \& {Cordes}}{{Schaffenroth}
  et~al.}{2014}]{schaffenroth14b}
{Schaffenroth} V.,  {Geier} S.,  {Heber} U.,  {Kupfer} T.,  {Ziegerer} E.,
  {Heuser} C.,  {Classen} L.,   {Cordes} O.,  2014, \mn@doi [\aap]
  {10.1051/0004-6361/201423377}, \href
  {https://ui.adsabs.harvard.edu/\#abs/2014A&A...564A..98S} {564, A98}

\bibitem[\protect\citeauthoryear{Schaffenroth, Barlow, Drechsel  \&
  Dunlap}{Schaffenroth et~al.}{2015}]{schaffenroth15}
Schaffenroth V.,  Barlow B.,  Drechsel H.,   Dunlap B.,  2015, A\&A, 576, 123

\bibitem[\protect\citeauthoryear{Schaffenroth et~al.,}{Schaffenroth
  et~al.}{2019}]{schaffenroth19}
Schaffenroth V.,  et~al., 2019, MNRAS, 630, 80

\bibitem[\protect\citeauthoryear{{Schaffenroth} et~al.,}{{Schaffenroth}
  et~al.}{2020}]{2020MNRAS.tmp.3451S}
{Schaffenroth} V.,  et~al., 2020, \mn@doi [\mnras] {10.1093/mnras/staa3661},
  \href {https://ui.adsabs.harvard.edu/abs/2020MNRAS.tmp.3451S} {}

\bibitem[\protect\citeauthoryear{{Schlafly}, {Meisner}  \& {Green}}{{Schlafly}
  et~al.}{2019}]{2019ApJS..240...30S}
{Schlafly} E.~F.,  {Meisner} A.~M.,   {Green} G.~M.,  2019, \mn@doi [\apjs]
  {10.3847/1538-4365/aafbea}, \href
  {https://ui.adsabs.harvard.edu/abs/2019ApJS..240...30S} {240, 30}

\bibitem[\protect\citeauthoryear{{Silvotti} et~al.,}{{Silvotti}
  et~al.}{2021}]{silvotti21}
{Silvotti} R.,  et~al., 2021, \mn@doi [\mnras] {10.1093/mnras/staa3332}, \href
  {https://ui.adsabs.harvard.edu/abs/2021MNRAS.500.2461S} {500, 2461}

\bibitem[\protect\citeauthoryear{Sterken}{Sterken}{2005}]{sterken05}
Sterken C.,  2005, in Sterken C.,  ed.,  Astronomical Society of the Pacific
  Conference Series Vol. 335, The Light-Time Effect in Astrophysics. p.~3

\bibitem[\protect\citeauthoryear{Vu{\v c}kovi{\'c}, {\O}stensen, Bloemen,
  Decoster  \& Aerts}{Vu{\v c}kovi{\'c} et~al.}{2008}]{vuckovic08}
Vu{\v c}kovi{\'c} M.,  {\O}stensen R.,  Bloemen S.,  Decoster I.,   Aerts C.,
  2008, in {Heber} U.,  {Jeffery} C.~S.,   {Napiwotzki} R.,  eds,  Astronomical
  Society of the Pacific Conference Series Vol. 392, Hot Subdwarf Stars and
  Related Objects. p.~199

\bibitem[\protect\citeauthoryear{Vu{\v c}kovi{\'c}, Bloemen  \&
  {\O}stensen}{Vu{\v c}kovi{\'c} et~al.}{2014}]{vuckovic14}
Vu{\v c}kovi{\'c} M.,  Bloemen S.,   {\O}stensen R.,  2014, in van Grootel V.,
  Green E.,  Fontaine G.,   Charpinet S.,  eds,  Meetings on Hot Subdwarf Stars
  and Related Objects Vol. 481, ASP Conference Series. p.~259

\bibitem[\protect\citeauthoryear{Vu{\v c}kovi{\'c}, {\O}stensen, N{\'e}meth,
  Bloemen  \& P{\'a}pics}{Vu{\v c}kovi{\'c} et~al.}{2016}]{vuckovic16}
Vu{\v c}kovi{\'c} M.,  {\O}stensen R.,  N{\'e}meth P.,  Bloemen S.,
  P{\'a}pics P.,  2016, A\&A, 586, 146

\bibitem[\protect\citeauthoryear{{Vu{\v c}kovi{\'c}}, {Aerts}, {{\O}stensen},
  {Nelemans}, {Hu}, {Jeffery}, {Dhillon}  \& {Marsh}}{{Vu{\v c}kovi{\'c}}
  et~al.}{2007}]{vuckovic07}
{Vu{\v c}kovi{\'c}} M.,  {Aerts} C.,  {{\O}stensen} R.,  {Nelemans} G.,  {Hu}
  H.,  {Jeffery} C.~S.,  {Dhillon} V.~S.,   {Marsh} T.~R.,  2007, \mn@doi
  [\aap] {10.1051/0004-6361:20077179}, \href
  {http://adsabs.harvard.edu/abs/2007A%26A...471..605V} {471, 605}

\bibitem[\protect\citeauthoryear{{Wolz} et~al.,}{{Wolz} et~al.}{2018}]{wolz18}
{Wolz} M.,  et~al., 2018, \mn@doi [Open Astronomy] {10.1515/astro-2018-0011},
  \href {https://ui.adsabs.harvard.edu/abs/2018OAst...27...80W} {27, 80}

\bibitem[\protect\citeauthoryear{Woudt et~al.,}{Woudt et~al.}{2006}]{woudt06}
Woudt P.,  et~al., 2006, MNRAS, 371, 1497

\makeatother
\end{thebibliography}


\label{lastpage}
\end{document}